\documentclass[10pt,journal,compsoc,twoside]{IEEEtran}
\ifCLASSOPTIONcompsoc
  \usepackage[nocompress]{cite}
\else
  \usepackage{cite}
\fi
\ifCLASSINFOpdf
\else
\fi
\usepackage{amsmath,amssymb,amsfonts}
\usepackage{graphicx}
\usepackage{textcomp}
\def\BibTeX{{\rm B\kern-.05em{\sc i\kern-.025em b}\kern-.08em
    T\kern-.1667em\lower.7ex\hbox{E}\kern-.125emX}}

\usepackage{hyperref}
\newcommand{\orcid}[1]{\href{https://orcid.org/#1}{\includegraphics[scale=0.7]{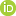}}}

\usepackage{comment}

\usepackage{algorithm}
\usepackage{algpseudocode}

\usepackage{xcolor, colortbl}
\usepackage{multirow}

\usepackage{soul}
\soulregister\ref7
\soulregister\cite7

\usepackage{float}
\usepackage{hhline}

\newcolumntype{C}[1]{>{\centering\arraybackslash}m{#1}}
\newdimen\NetTableWidth
 \NetTableWidth=\dimexpr
    \linewidth - 8\tabcolsep - 5\arrayrulewidth

\definecolor{tabhead}{RGB}{220,220,220}
\definecolor{classcol}{RGB}{169,169,169}

\makeatletter
\newcommand{\StatexIndent}[1][3]{%
  \setlength\@tempdima{\algorithmicindent}%
  \Statex\hskip\dimexpr#1\@tempdima\relax}
\algdef{S}[WHILE]{WhileNoDo}[1]{\algorithmicwhile\ #1}%
\makeatother

\begin{document}
%
\title{Leveraging Siamese Networks for One-Shot Intrusion Detection Model}

%
%
%
%

\author{
Hanan~Hindy\orcid{0000-0002-5195-8193},~\IEEEmembership{Student Member,~IEEE},
Christos~Tachtatzis\orcid{0000-0001-9150-6805},~\IEEEmembership{Senior Member,~IEEE}, 
Robert~Atkinson\orcid{0000-0002-6206-2229},~\IEEEmembership{Senior Member,~IEEE},
David~Brosset\orcid{0000-0002-9677-1445}, 
Miroslav~Bures\orcid{0000-0002-2994-7826}, 
Ivan~Andonovic\orcid{0000-0001-9093-5245},~\IEEEmembership{Senior Member,~IEEE},
Craig~Michie\orcid{0000-0001-5132-4572}, and Xavier~Bellekens\orcid{0000-0003-1849-5788},~\IEEEmembership{Member,~IEEE}
\thanks{This work has been submitted to the IEEE for possible publication. Copyright may be transferred without notice, after which this version may no longer be accessible.}
}

\markboth{Journal of \LaTeX\ Class Files,~Vol.~14, No.~8, August~2015}%
{H. Hindy \MakeLowercase{\textit{et al.}}: Siamese Networks for Intrusion Detection}
%



\IEEEtitleabstractindextext{%
\begin{abstract}
The use of supervised Machine Learning~(ML) to enhance Intrusion Detection Systems has been the subject of significant research. Supervised ML is based upon learning by example, demanding significant volumes of representative instances for effective training and the need to re-train the model for every unseen cyber-attack class. However, retraining the models in-situ renders the network susceptible to attacks owing to the time-window required to acquire a sufficient volume of data. Although anomaly detection systems provide a coarse-grained defence against unseen attacks, these approaches are significantly less accurate and suffer from high false-positive rates.  Here, a complementary approach referred to as `One-Shot Learning', whereby a limited number of examples of a new attack-class is used to identify a new attack-class (out of many) is detailed. The model grants a new cyber-attack classification without retraining. A Siamese Network is trained to differentiate between classes based on pairs similarities, rather than features, allowing to identify new and previously unseen attacks. The performance of a pre-trained model to classify attack-classes based only on one example is evaluated using three datasets.  Results confirm the adaptability of the model in classifying unseen attacks and the trade-off between performance and the need for distinctive class representation.

\end{abstract}

\begin{IEEEkeywords}
Artificial Neural Network,
Continuous Learning,
CICIDS2017,
Intrusion Detection,
KDD Cup'99,
NSL-KDD,
Machine Learning, 
One-Shot Learning,
Siamese Network.
\end{IEEEkeywords}}

\maketitle

\IEEEdisplaynontitleabstractindextext

%
\IEEEpeerreviewmaketitle

\IEEEraisesectionheading{\section{Introduction}}
\label{sec:introduction}

\IEEEPARstart{I}ntrusion Detection System (IDS) development has its roots in statistical models~\cite{patcha2007overview}, and has recently evolved to the use of Machine Learning (ML)~\cite{7307098} based on hybrid models and adaptive techniques~\cite{HananHindy2018}. Developments to date have highlighted two fundamental considerations in the design of effective supervised ML-based IDS; (a)~availability of a large and representative historian of cyber-attacks consisting of many thousands of instances \cite{RefWorks:doc:5c7e5b20e4b0f05d47246237} and (b)~the time window resulting from the need to retrain models after the emergence of a new attack class has been recorded, renders the network open to damaging attacks. Supervised ML models are very accurate at identifying cyber-attacks previously been trained to recognise, but significantly under-perform for new unseen and `zero-day' attacks that emerge. Anomaly detection approaches have been explored to address the issue and whilst these schemes provide better performance against unseen attacks, their efficacy is inferior against known attacks when compared to supervised ML approaches. Further, anomaly-based approaches are also limited under multiple new attacks scenarios as they are simply classified into the same anomalous group, in so doing restricting the range of attack-specific countermeasures that can be employed.

Here, the development and evaluation of an ML-enabled approach that provides improved attack identification in the period between a range of previously unseen attacks at onset is reported and the deployment of a robust supervised ML model that informs on the most effective countermeasures. The methodology - referred to as One-Shot Learning - centres on the use of a Siamese Network, shown to be effective in identifying new classes based on one (or only a few) examples of a new class. An alternative approach is to create synthetic examples based on the domain knowledge of new attacks; however, this is challenging requiring a considerable amount of time to replicate a suitable representation of an environment with appropriate parameters, and is consequently subject to human error owing to cognitive biases. 

One-Shot Learning was inspired by the generalisation learning ability of human beings. As discussed by Vinyals~\textit{et al.}~\cite{RefWorks:doc:5d4834f3e4b04c1c1a583763}, ``Humans learn new concepts with very little supervision, yet our best deep learning systems need hundreds or thousands of examples''~\cite{RefWorks:doc:5d4834f3e4b04c1c1a583763}. Therefore, One-Shot learning models aim at classifying previously unseen classes using one instance. The idea is to rely on previously seen classes and learn patterns and similarities instead of fitting the ML model to fixed classes. Few-Shot (N-Shot) learning is similar to One-Shot learning with a flexibility of using a few (N) instances to classify a class instead of one~\cite{sun2019meta}.



A Siamese Network is a network composed of two ``twin'' networks that are trained simultaneously to learn the similarity of two instances called a pair. Leveraging this similarity-based learning, a previously unseen class could be added to the network without retraining.
The initial stage of the development is the training phase. The Siamese Network is trained using similarities that discriminate between $K$ classes; benign traffic and the $K-1$ classes of known cyber-attacks. Any new traffic instance $P$ is then compared against all known classes (used during training) plus an additional class ($K+1$ classes) where only a limited number of examples of class `$K+1$' are available, such as might be the case on the appearance of a new cyber-attack. This is achieved without any form of additional training. 

\vspace{2mm}
The contributions of the paper are; (a)~the use of a Siamese Network model to successfully classify cyber-attacks based on pair similarities, not proposed for Cyber Security usage to date. (b)~ evaluation of the proposed model performance to detect a new cyber-attack class based on one labelled instance without re-training. (c)~comparison of the impact of a few labelled instances of the new attack class on detection performance.

The remainder of the paper is organised as follows; Section~\ref{sec:background} details the main features of Siamese Networks; 
Section~\ref{sec:one-shot} presents the methodology governing the training of the Siamese Network and its evaluation is explained showing the potential of the network to identify a new attack class based on a few (previously collected and labelled) examples of that attack class without retraining.
Section~\ref{sec:datasets} presents the properties of the data sets and their corresponding attack classes used in model development and performance evaluation; 
; the performance of the model is assessed in Section~\ref{sec:one-shot-evaluation}; 
conclusions are drawn in Section~\ref{sec:conclusion}.




\section{Background}
\label{sec:background}
In supervised machine learning, a relationship exists between model complexity and the volume of training data; too few training examples and the model will over-fit, resulting in an unnecessarily complex model that produces poor results. Therefore, securing sufficient and representative data is a limiting factor in model development and performance~\cite{RefWorks:doc:5bf40cd2e4b0c53ba57ac274}. In practice, accessing and/or generating sufficiently large and representative training examples is a complex challenge and may involve significant manual effort and processing time~\cite{roh2018survey}. Nonetheless, there are  publicly available data sets for training IDS systems, notably the CICIDS2017 and the NSL-KDD sets. These data are used to pre-train the Siamese Network, subsequently, in the evaluation of the performance of the model in identifying a new class of attack after a limited number of that class' samples has been recorded.


An alternative approach is to utilise `\textit{Transfer Learning}' to mitigate the need for large volumes of training data~\cite{pan2010survey, Weiss2016}. The premise of Transfer Learning to solve the target problem $T$ (where data are limited), is to create a model $M$ for a similar problem $T^\prime$ where large amounts of data are readily available. The initial model $M$ is then `transferred' to the target problem $T$ and partially re-trained on the small data set. The rationale is that the initial training on $T^\prime$, yields training weights which discover features useful for the problem domain and hence applicable to the target problem $T$; hence after retraining, the model learns and generalises faster on the small data set~\cite{RefWorks:doc:5c88f2c4e4b0078bb5c79834}. Transfer Learning is a common approach in the image processing domain~\cite{8241712} where for example, models are trained on the ImageNet data set~\cite{nguyen2018deep, russakovsky2015imagenet, ngiam2018domain}. Despite the potential of Transfer Learning as a viable solution, it does not eliminate the need for retraining. 


\vspace{2mm}
One-Shot learning, first reported  by Li Fei-Fei~\textit{et al.}~\cite{RefWorks:doc:5bf40968e4b08f3b86fda7ba}, is inspired by human generalisation learning and has been applied in multiple domains with the most prominent being image and video processing~\cite{RefWorks:doc:5bf4200fe4b0d4880bbf46e9, RefWorks:doc:5bf41f66e4b045abd39974a0, RefWorks:doc:5bf41f5fe4b0f174e717833a}. It has also been used in other domains, such as robotics~\cite{RefWorks:doc:5bf41fc6e4b03fd0c4898245}, language processing ~\cite{RefWorks:doc:5bf4200fe4b0c53ba57ac7c4, RefWorks:doc:5bfc0cd2e4b09e17fc6e84c6} and drug discovery~\cite{RefWorks:doc:5bf4156ae4b06df66522e3f3}.
Based on the literature, the Siamese Network is the most frequently used. Various architectures have been proposed and assessed as the building block for the twin network (i.e., CNN~\cite{RefWorks:doc:5d483701e4b092b0fda69856, RefWorks:doc:5d4836f3e4b03defc5364238}, RNN~\cite{RefWorks:doc:5d483798e4b0848866848217} and GNN~\cite{garcia2017few}). Matching Networks~\cite{RefWorks:doc:5d4834f3e4b04c1c1a583763}, Prototypical Networks~\cite{RefWorks:doc:5d482ff2e4b0e3b2b850572c}, Imitation Learning~\cite{RefWorks:doc:5d483518e4b0e3b2b850588a} and Autoencoders~\cite{7124463}, particularly in the image processing domain, but amenable to be generalised to other domains. To the best of the authors' knowledge, the development reported here is the first proposing a One-Shot IDS model implementation.



\subsection{Siamese Network Architecture}
\label{subsec:siamese_network_architecture}
Siamese Networks were first introduced by Bromley~\textit{et al.}~\cite{RefWorks:doc:5bfbe8c5e4b03539f0f7ca0d} in the 90s to solve the problem of matching hand-written signatures, subsequently adapted to other domains. Popular implementations of Siamese Networks for image and video processing are presented by Koch~\textit{et al.}~\cite{koch2015siamese}, Yao~\textit{et al.}~\cite{RefWorks:doc:5bfbf527e4b0bf300b6688b1} and Varior~\textit{et al.}~\cite{RefWorks:doc:5bfbf41de4b0c12eaacd420e}. Moreover, it has been implemented for Natural Language Processing~(NLP) tasks~\cite{benajiba2019siamese, zhu2018dependency} and for the retrieval of similar questions~\cite{das2016together}.

\begin{figure}[tbh]
    \centering
    \includegraphics[width=0.8\columnwidth]{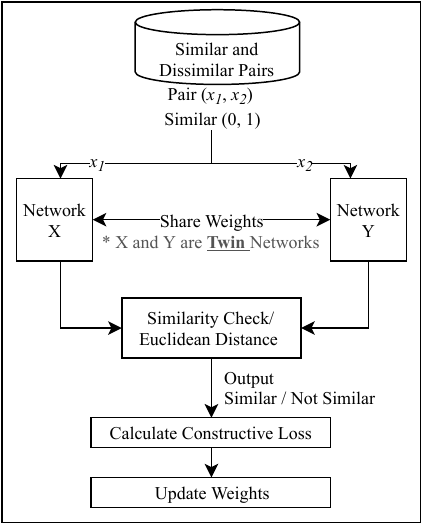}
    \caption{Siamese Network Architecture.}
    \label{fig:siamese-netowrk}
\end{figure}

Figure~\ref{fig:siamese-netowrk} depicts the Siamese network architecture. As shown, the network is composed of two identical sub-networks that share weights. Twin networks pass their output to a similarity module, which in turn is responsible for calculating the distance defining ``how alike'' the two inputs are. The output is compared to the given similarity (i.e. whether or not the pair are similar), the loss is calculated, and the weights are then adjusted.

\vspace{3mm}
Formally~\cite{koch2015siamese, SHAHAM201852}, given a pair of inputs $(x_1, x_2)$ and a twin network $(X, Y)$, such that $x_1$ is the input of $X$ and $x_2$ is the input of $Y$, the similarity can be computed using 
Euclidean distance (equation~\ref{eq:similarity_2}):


\begin{equation}
\label{eq:similarity_2}
    d = \mid\mid f_1(x_1) - f_2(x_2) \mid\mid_2
\end{equation}
such that $f_1$ and $f_2$ are the outputs of Networks $X$ and $Y$ respectively $f_1 \equiv f_2$ since $X$ and $Y$ are twin networks. Ultimately, the training goal is to minimise the overall loss $l$ as defined in Equation~\ref{eq:loss}; for each given batch $i$ of input pairs $(x_1, x_2)_i$ and label vector $y_i$, such that $y_i(x_1, x_2)_i = 1$ if $x_1$ and $x2$ belong to the same class and $0$ otherwise. 

\begin{equation}
\label{eq:loss}
\begin{split}
    l(x_1, x_2)_i = y(x_1, x_2)_i \log d_i +
    (1 - y(x_1, x_2)_i) \log (1-d_i) \\ + \lambda \mid w \mid*2
\end{split}
\end{equation}
\noindent
such that $\lambda$ is a $l_2$ regularisation parameter. 

However, the loss function is sensitive to outliers (i.e. dissimilar pairs with large distances) which disproportionately affect the gradient estimation. An alternative loss function is the constructive loss shown in equation~\ref{eq:constructive_loss} proposed by Chopra, Hadsell and LeCun~\cite{RefWorks:doc:5d03af8ee4b0365aad08d9d1, RefWorks:doc:5d03af26e4b004e1aa1188b4}. The constructive loss caps the contribution of dissimilar pairs if the distance is within a specified margin $m$~\cite{RefWorks:doc:5d03af26e4b004e1aa1188b4}, hence limiting the effect of large distances.

\begin{equation}
\label{eq:constructive_loss}
\begin{split}
    l(x_1, x_2) = \sum_{n=1}^{B} &y(x_1, x_2)_i * (d_i)^2 \\ &+ (1-y(x_1, x_2)_i) * (max( m -d_i, 0))^2
\end{split}
\end{equation}
\noindent
such that $m > 0$ is a margin. In this study, the margin was set to $m=1$~\cite{RefWorks:doc:5d03af26e4b004e1aa1188b4}. 

After training, given any two pairs, the network is capable of calculating their degree of similarity, $d_i \in [0,1]$, $d_i$ mirror the degree of similarity for the pair; the lower the $d_i$, the closer the pair. Batches of pairs are used to train the network. Note, however, that an equal number of similar and dissimilar pairs are used in the batch.


Here, Feed-forward Artificial Neural Networks~(ANN) are used as the building block of the twin network. The details of the architecture (i.e., the number of layers, neurons, etc.) are provided in Section~\ref{sec:one-shot}.



\section{Siamese Network Model}
\label{sec:one-shot}
In this section, the proposed Siamese Network model is used as the One-Shot learning architecture. The performance of the network on classifying a new cyber-attack class without the need to retrain is evaluated with the new attack class represented by a limited number of labelled samples. 

\begin{figure}[tb]
    \centering
    \includegraphics[width=\columnwidth]{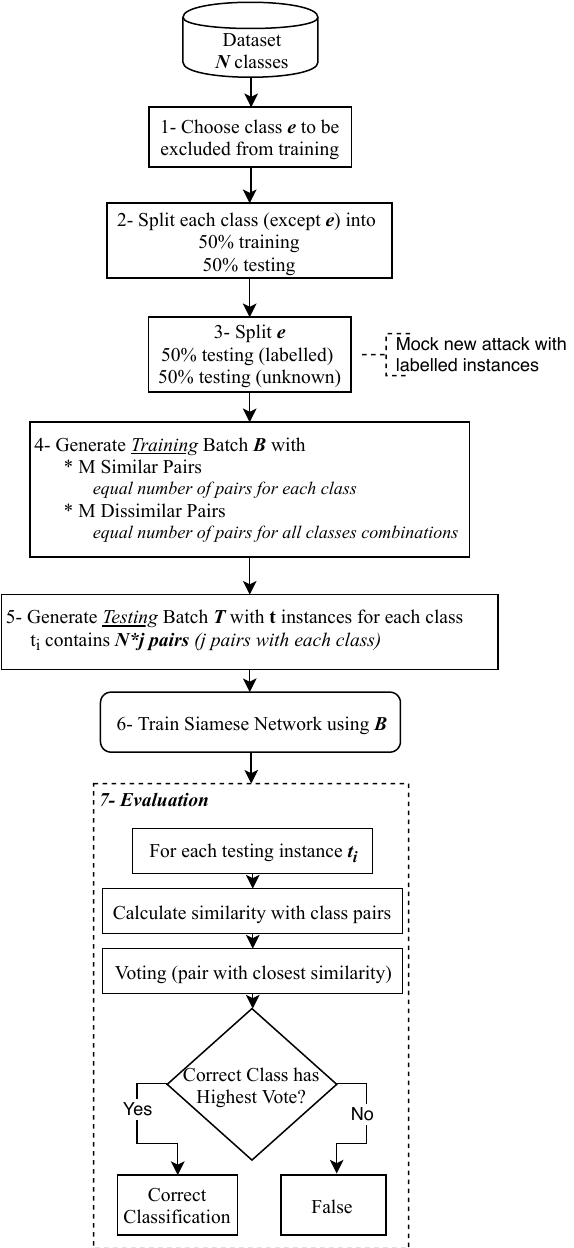}
    \caption{Siamese Network for Intrusion Detection System (One-Shot).}
    \label{fig:one_shot_architecture}
\end{figure}
   
Figure~\ref{fig:one_shot_architecture} shows the process of establishing the intrusion detection model based on one-shot learning and illustrates the methodology of assessing performance for new attack classes without retraining the model.

Given a data set with $N$ classes, first, an attack class $e$ is chosen to act as the new cyber-attack; this class is excluded from the training process~(Figure~\ref{fig:one_shot_architecture}-(1)). 
Second, for the remaining $K$ classes after excluding $e$ ($N-1$ classes), each class instances are split into two, as shown in Figure~\ref{fig:one_shot_architecture}-(2). 
Collectively, the first `half' is used as a pool of instances to generate the training set pairs both similar and dissimilar, as shown in Figure~\ref{fig:one_shot_architecture}-(4); the second `half' is used as the evaluation pool of instances. 

Class $e$ is used to mimic a real-life situation in which a new attack is detected with only a few labelled samples available. Therefore, the instances of $e$ are split in two halves (Figure~\ref{fig:one_shot_architecture}-(3)), the first half representing a pool of labelled and the second half a pool of unlabelled (new) instances. 

Since the model relies on random pair generation, pairs are drawn out randomly from the pools of instances. The rational for having pools of instances and to draw out pairs randomly is to hinder any selection bias either during training (i.e. selecting similar and dissimilar pairs) or during evaluation of the new class (i.e. selecting the labelled instances that best represent this class). Furthermore, the uniqueness of the pairs - no duplicates - is ensured. A ``set'' data structure is used. it is added to the batch of pairs unless that pair is already contained within the set. This is demonstrated in Algorithm~\ref{alg:batch}.

During evaluation, an instance $i$ is paired with one random instance from each class. The instances are drawn out of the pool of testing instances, resulting in $N$ pairs. The similarity is then calculated for the $N$ pairs. Instance $i$ is classified (labelled) based on the pair with the highest similarity (i.e. least distance). 

As discussed in Section~\ref{sec:one-shot-evaluation}, to determine the trade-off between the number of labelled instances of the new attack class and accuracy, the process is repeated $j$ times for each instance $i$. Majority voting is then applied to deduce the instance label; the class with the highest votes is used as instance $i$ label (Figure~\ref{fig:one_shot_architecture}-(7)).

\begin{algorithm}[b]
\caption{Train and Test Siamese Network}
\label{alg:main-one-shot}
\hspace*{\algorithmicindent} \textbf{Input:} Attacks Dataset \\
\hspace*{\algorithmicindent} \textbf{Output:} Trained Siamese Network Evaluation
\begin{algorithmic}[1]
\Ensure $dataset = \{c_1, c_2,\ldots, c_n\} \: s.th. \: n\geq 3$
\State ${train\_batch\_size, test\_batch\_size} \leftarrow 30,000$ 
\State $n\_epochs \leftarrow 2000$
\State $excluded\_class = $ random class $e$ s.th. \: $e \in dataset$
\State $training\_classes = dataset - e$
\State $training = 50\%\: c_i \:\forall c_i \in training\_classes$
\State $testing = dataset\cap \overline{training}$
\State $batch \leftarrow$ \Call{GetTrainingBatch}{$train\_batch\_size$}

\State Build Siamese Network with Random Weights

\For{$i=0$ to $n\_iterations$}
\State Update Siamese Network Weights based on $batch$ 
\EndFor

\State \Call{Evaluate}{test\_batch\_size}
\end{algorithmic}
\end{algorithm}

\begin{algorithm}[h]
    \caption{Generate Training Batch}
    \label{alg:batch}
    \hspace*{\algorithmicindent} \textbf{Input:} Dataset of $K (N-1)$ classes, Batch Size \\
    \hspace*{\algorithmicindent} \textbf{Output:} Batch of similar and dissimilar pairs \\ 
    \hspace*{\algorithmicindent} and associated labels (0: dissimilar, 1: similar)
    \begin{algorithmic}[1]
    \Function{GetTrainingBatch}{batch\_size}
        \State $num\_similar\_pairs = batch\_size / 2 $
        \State $num\_dissimilar\_pairs = batch\_size / 2 $
        \State $num\_similar\_pairs\_per\_class $\StatexIndent[3] $=num\_similar\_pairs / K $
        \State $all\_combinations = combinations(K)$
        \State $num\_dissimilar\_pairs\_per\_combination$\StatexIndent[3]$ = num\_dissimilar\_pairs /len(all\_combinations) $
        \State $pairs\_set \leftarrow \{\}$
        \For{$c$ in $K$}
            \For{$i=0$ to $num\_similar\_pairs\_per\_class$}
                \State ($ins_1, ins_2) \leftarrow$ 2 random instances $\in c_i$
                \If {($ins_1, ins_2) \in pairs\_set$}
                \State \textbf{go to} 10
                \EndIf
                \State $pairs[i] \leftarrow \{ins_1, ins_2\}$
                \State $pairs\_set.add(\{ins_1, ins_2\})$
            \EndFor
        \EndFor
        
        \For{$c_1, c_2$ in $all\_combinations$}
            \For{$i=0$ to \StatexIndent[3] $num\_dissimilar\_pairs\_per\_combination$}
                \State $ins_1 \leftarrow$ random instance $\in c_1$
                \State $ins_2 \leftarrow$ random instance $\in c_2$
                \If {($ins_1, ins_2) \in pairs\_set$}
                \State \textbf{go to} 20
                \EndIf
                \State $pairs[i] \leftarrow \{ins_1, ins_2\}$
                \State $pairs\_set.add(\{ins_1, ins_2\})$
            \EndFor
        \EndFor
        
        \State $targets[0..batch\_size/2] \leftarrow 0 $ \Comment{Similar}
        \State $targets[batch\_size/2..batch\_size] \leftarrow 1 $ \Comment{Dissimilar}
        
        \State \textbf{return} $pairs$, $targets$
    \EndFunction
    \end{algorithmic}
\end{algorithm}

\vspace{2mm}
Algorithm~\ref{alg:main-one-shot} summarises the overall process of training and testing the model. Initially, the data set is split as shown in Figure~\ref{fig:one_shot_architecture}. The model is trained for a specified number of epochs with the generated batch of pairs as described in Algorithm~\ref{alg:batch}. 
The $batch\_size = 30,000$ is based on the literature recommendation for the advisable Siamese Network training batch size~\cite{8788519, koch2015siamese, Serra2015ICCV}. It is important to note that the classes are equally represented in both the training and testing batches. Note that the data set should have at least 3 classes, otherwise, the model converges to a 50\% similarity output and fails to train adequately. Algorithm~\ref{alg:batch} shows the training batch generation process.

An equal number of instances are used from each class for evaluation (Algorithm~\ref{alg:classification}). For each new instance, a pair is selected with each class using the new instance and a random instance from each class. The similarity is calculated for each pair. The pair with the closest similarity contributes to the classification result. The process is performed $j$ times and majority voting is used to collate the results ($j \in {1, 5, 10, 15, 20, 25, 30}$). For class $e$ (the attack class that is excluded from training), the first half acts as the pool of labelled and the second half act as the pool of new unlabelled instances.

\begin{algorithm}[tb]
    \hspace*{\algorithmicindent} \textbf{Input:} Trained Siamese Network, Batch Size, Excluded Class ($e$) \\
    \hspace*{\algorithmicindent} \textbf{Output:} Accuracy
    \caption{Evaluate Model}
    \label{alg:classification}
    \begin{algorithmic}[1]
    \Function{Evaluate}{batch\_size}
        \State $n\_correct \leftarrow 0$
        \State $num\_per\_class \leftarrow batch\_size / N$
        
        \For{$c$ in $N$}
            \For{$i=0$ to $num\_per\_class$}
                \For{$j=0$ to $5$}
                    \If{$c == e$}
                        \State $ins_1 \leftarrow$ \StatexIndent[6] random instance $\in c\_testing$
                    \Else
                        \State $ins_1 \leftarrow$ \StatexIndent[6] random instance  $\in e\_unlabelled$
                    \EndIf
                    \State $pairs \leftarrow(ins_1,$ \StatexIndent[6] random instance $\in x \forall x \in K)$
                    \State $pairs.append(ins_1,$ \StatexIndent[6] random instance $\in e\_labelled$
                    
                   \State $similarities \leftarrow model.predict(pairs)$
                    \State $votes[argmin(similarities)] += 1$
                \EndFor
                \If{$argmax(votes) == c $}
                \State $n\_correct += n\_correct + 1$
                \EndIf
                \State $confusion\_matrix[c, argmax(votes)] += 1$
            \EndFor
        \EndFor
        \State $accuracy = n\_correct * 100/batch\_size$
        \State \textbf{return} $accuracy, confusion\_matrix$
    \EndFunction
    \end{algorithmic}
\end{algorithm}

\vspace{2mm}
The model evaluation yields a Confusion Matrix~(CM) that visualises the performance. A sample CM is presented in Table~\ref{tab:sample_CM}. Each row of the CM represents a class; True Positive~(TP) is the number of attack instances correctly classified as attack; True Negative~(TN) is the number of normal instances correctly classified as normal; False Positive~(FP) is the number of normal instances wrongly classified as attack; False Negative~(FN) is the number of attack instances wrongly classified as normal.

\begin{table}[h]
    \centering
    \caption{Sample Confusion Matrix}
    \renewcommand{\arraystretch}{1.2}
    \begin{tabular}{|C{0.15\NetTableWidth}|C{0.125\NetTableWidth}|C{0.125\NetTableWidth}|C{0.125\columnwidth}|C{0.125\NetTableWidth}|C{0.125\NetTableWidth}|C{0.125\NetTableWidth}}
    \hhline{~-----~}
    \multicolumn{1}{c|}{} &
    \multicolumn{5}{c|}{\cellcolor{tabhead}Predicted Class}  \\
    \hline
    \rowcolor{tabhead}
     Correct &  Normal & Attack\textsubscript1 & Attack\textsubscript2 & Attack\textsubscript3 & Attack\textsubscript4 \\ \hline
    \cellcolor{tabhead}Normal & \cellcolor{tabhead}TN & FP\textsubscript1 & FP\textsubscript2 & FP\textsubscript3 & FP\textsubscript4 \\ \hline
    \cellcolor{tabhead}Attack\textsubscript1 & FN\textsubscript1 & \cellcolor{tabhead}TP\textsubscript{11} & TP\textsubscript{12} & TP\textsubscript{13} & TP\textsubscript{14} \\\hline
    \cellcolor{tabhead}Attack\textsubscript2 & FN\textsubscript2 & TP\textsubscript{21} & \cellcolor{tabhead} TP\textsubscript{22} & TP\textsubscript{23} & TP\textsubscript{24} \\\hline
    \cellcolor{tabhead}Attack\textsubscript3 & FN\textsubscript3 & TP\textsubscript{31} & TP\textsubscript{32} &  \cellcolor{tabhead} TP\textsubscript{33} & TP\textsubscript{34} \\\hline
    \cellcolor{tabhead}Attack\textsubscript4 & FN\textsubscript4 & TP\textsubscript{41} & TP\textsubscript{42} &   TP\textsubscript{43} & \cellcolor{tabhead} TP\textsubscript{44} \\\hline
    \end{tabular}

    \label{tab:sample_CM}
\end{table}

\noindent
The overall accuracy is calculated as shown in Equation~\ref{eq:overall_accuracy}. True Positive Rate~(TPR) and False Negative Rate~(FPR) for each class are shown in Equation~\ref{eq:tpr} and Equation~\ref{eq:fnr} respectively; finally, True Negative Rate~(TNR) and False Positive Rate~(FPR) are calculated using Equation~\ref{eq:tnr} and Equation~\ref{eq:fpr} respectively.

\begin{equation}
\label{eq:overall_accuracy}
\begin{aligned}
    &Overall Accuracy =\\& \frac{TN + \sum_{i=1}^{4}TP_{ii}} {TN + \sum_{i=1}^{4}\sum_{j=1}^{4}TP_{ij} + \sum_{i=1}^{4}FP_{i}  + \sum_{i=1}^{4}FN_{i} }
\end{aligned}
\end{equation}

\begin{equation}
\label{eq:tpr}
    TPR_i = \frac{TP_{ii}} {FN_i + \sum_{j=1}^{4}TP_{ij}}
\end{equation}

\begin{equation}
\label{eq:fnr}
    FNR_i = \frac{FN_{i}} {FN_i + \sum_{j=1}^{4}TP_{ij}}
\end{equation}

\begin{equation}
\label{eq:tnr}
    TNR = \frac{TN} {TN + \sum_{i=1}^{4}FP_{i}}
\end{equation}

\begin{equation}
\label{eq:fpr}
    FPR = \frac{\sum_{i=1}^{4}FP_{i}}{TN +\sum_{i=1}^{4}FP_{i}}
\end{equation}
\section{Datasets}
\label{sec:datasets}
Three data sets are used to evaluate the proposed models; 
two benchmark IDS data sets, specifically, CICIDS2017 and NSL-KDD and KDD Cup'99. The latter is used in comparison to the NSL-KDD to demonstrate the effectiveness of clean data when generating training pairs and also, when introducing new attacks to the trained model.

Each data set contains $N$ classes. $K$  classes are used to train the network, such that $K = N - 1$. The $K$ classes include normal/benign and $K-1$ attack classes. 
The instances of each of the $K$ class act as a pool used to generate similar and dissimilar pairs. 
Furthermore, one class is used to simulate a new attack, mimicking the situations in which little/limited data is available for a new attack. The pair generation details and the experiments are further discussed in Section~\ref{sec:one-shot}.

An overview of each data set is presented in the following subsections. 

\subsection{CICIDSS2017}
CICIDS2017 ~\cite{RefWorks:doc:5d4acfdae4b002b95f900284} is a recent data set generated by the Canadian Institute for Cyber-security~(CIC) comprising up-to-date benign, insider and outsider attacks. Traffic flows were generated and labelled using the provided `.pcap' files. Table~\ref{tab:cicattacks} lists the attacks used and the number of instances/flows for each.

\begin{table}[thb]
    \centering
    \caption{CICIDS Classes and Corresponding Number of Occurrences (1)}
    \renewcommand{\arraystretch}{1.3}
    \begin{tabular}{|c|c|c|}
         \hline
         \rowcolor{gray!30}
         & Class & \# of Occurrences\\ \hline
         1 & Normal & 248607 (90.50\%) \\ \hline
         2 & DoS (Hulk) & 14427 (5.25\%) \\ \hline
         3 & DoS (Slowloris) & 2840 (1.03\%) \\ \hline
         4 & FTP Brute Force & 5228 (1.9\%) \\ \hline
         5 & SSH Brute Force & 3627 (1.32\%) \\ \hline
    \end{tabular}
    \label{tab:cicattacks}
\end{table}

\subsection{KDD Cup'99}
The KDD Cup'99~\cite{RefWorks:doc:5b06959ce4b078503013ca8c}, although old, is still considered as the classic benchmark data set used in the evaluation of IDS performance. More than 60\% of the research in the past decade (2008 - 2018) has been evaluated using KDD'99~\cite{HananHindy2018}. KDD Cup'99 covers 4 attack classes alongside normal activity. The attacks contained in the data set are; Denial of Service~(DoS), Root to Local~(R2L), User to Root~(U2R) and probing.

The KDD Cup'99 data set is relatively large, however, the provider has made available a reduced subset of \textasciitilde 10\%~\cite{KDDCup1916:online}. For the purposes of evaluation here, only the smaller subset is used. 
Table~\ref{tab:kddinstances} shows the number of instances per class for the KDD Cup'99 data set. 

\begin{table}[thb]
    \centering
    \caption{KDD Cup'99 Classes and Corresponding Number of Occurrences}
    \renewcommand{\arraystretch}{1.3}
    \begin{tabular}{|c|c|c|}
         \hline
         \rowcolor{gray!30}
         & Class & \# of Occurrences\\ \hline
         1 & Normal & 97278 (19.70\%) \\ \hline
         2 & DoS & 391458 (79.24\%) \\ \hline
         3 & Probe & 4107 (0.82\%) \\ \hline
         4 & U2R & 1128 (0.23\%) \\ \hline
         5 & R2L & 52 (0.01\%) \\ \hline
    \end{tabular}
    \label{tab:kddinstances}
\end{table}

\subsection{NSL-KDD}
The NSL-KDD~\cite{RefWorks:doc:5b227bd4e4b0d1cffc0657b0} data set was proposed by the CIC to overcome the problems of the KDD Cup'99 set discussed by Tavallaee \textit{et al.}~\cite{RefWorks:doc:5d03bf33e4b0d6a9f11ef38c}.
Similar to KDD Cup'99, NSL-KDD covers 4 attack classes alongside normal activity. NSL-KDD is used for evaluating the effect of enhancing and filtering a data set on the similarity learning and performance. 
Table~\ref{tab:nsl-kddinstances} shows the number of instances per class for the NSL-KDD data set. 

\begin{table}[thb]
    \centering
    \caption{NSL-KDD Classes and Corresponding Number of Occurrences}
    \renewcommand{\arraystretch}{1.3}
    \begin{tabular}{|c|c|c|}
         \hline
         \rowcolor{gray!30}
         & Class & \# of Occurrences \\ \hline
         1 & Normal & 67343 (53.46\%) \\ \hline
         2 & DoS & 45927 (36.47\%) \\ \hline
         3 & Probe & 11656 (9.25\%) \\ \hline
         4 & U2R & 995 (0.78\%) \\ \hline
         5 & R2L & 52 (0.04\%) \\ \hline
    \end{tabular}
    \label{tab:nsl-kddinstances}
\end{table}

\noindent
NSL-KDD and KDD Cup'99 data sets have already been pre-processed and 42 features extracted, a total of 118 features after encoding the categorical features. For the CICIDS2017, 31 bidirectional flow features are extracted. It is worth noting that no feature engineering or selection is performed to ensure that the excluded class from training does not indirectly influence the feature set.

Recent surveys examined the use of ML for IDS~\cite{10.1007/978-981-13-1921-1_35}. Furthermore, Thomas and Pavithran~\cite{8649498} study the recent ML techniques evaluated using the NSL-KDD data set. While, Panwar \textit{et al.}~\cite{singh2019evaluation} evaluate the usage of ML on CICIDS-2017 data set. Although there are various manuscript using ML for IDS, comparing the proposed model with recent IDS models is not applicable. This is because the proposed model leverages One-Shot learning, therefore, it cannot be in comparison with classical classification models.
\section{One-Shot Evaluation}
\label{sec:one-shot-evaluation}
\renewcommand{\arraystretch}{1.35}

The evaluation specifies how accurately the proposed network can classify both classes used in training and new attack classes without the need for retraining. The model leverages similarity-based learning. The new attack class is represented using one sample to mimic the labelling process of new attacks. 

For each data set evaluation, multiple experiments are conducted. Specifically, $K$ ($N - 1$) experiments are evaluated, where $N$ is the number of classes and $K$ is the number of attack classes in order to evaluate the performance of the Siamese Network when using a different set of attack classes for training and evaluation. In each experiment, a separate attack class ($e$) is excluded, one at a time. The CM is presented alongside the overall model accuracy for each experiment. 

The results of the evaluation of the performance impact of the number of labelled samples ($j$) of the new attack class $e$ are presented in terms of  overall accuracy, new attack True Positive Rate~(TPR) and False Negative Rates~(FNR), Normal True Negative Rate~(TNR) and False Positive Rate~(FPR), listed using $j$ instances for majority voting, where $j \in {1, 5, 10, 15, 20, 25, 30}$. The CMs use $j = 5$.

First, the CMs of the CICIDS2017 One-Shot, excluding SSH class is presented in Table~\ref{tab:cicids-ssh-one-shot} and excluding FTP in Table~\ref{tab:cicids-ftp-one-shot}. The overall accuracy is 81.28\% and 82.5\% respectively. The results demonstrate the network capability to adapt to the emergence of a new cyber-attack after training. It is important to note that the new attack class performance is 73.03\% and 70.03\% for SSH and FTP respectively. 
Moreover, the added class demonstrates low FNRs, specifically 8\% and 15\% for FTP and SSH respectively. On inspection of Table~\ref{tab:cicids-ssh-one-shot-pairs} and Table~\ref{tab:cicids-ftp-one-shot-pairs}, it is evident that using five labelled instances of the new attack class results in an increase in both the overall accuracy and the TPR together with a drop in the FNR. Using only 1 labelled instance demonstrates a comparably poorer performance owing to the instance selection randomness, which could result in either a good or a bad class representative. However, using 5 random labelled instances boosts  performance, reinforcing the importance of having distinctive class representatives.

The remainder of the CICIDS2017 performance evaluation results are characterised by similar behaviour and are listed as follows. DoS (Hulk) results are presented in Table~\ref{tab:cicids-hulk-one-shot} and Table~\ref{tab:cicids-hulk-one-shot-pairs}, while DoS (Slowloris) in Table~\ref{tab:cicids-slowloris-one-shot} and Table~\ref{tab:cicids-slowloris-one-shot-pairs}. 

\begin{table}[bht]
    \centering
    \caption{CICIDS2017 One-Shot Confusion Matrix (SSH not in Training)}
    \begin{tabular}{|C{0.15\NetTableWidth}|C{0.125\NetTableWidth}|C{0.125\NetTableWidth}|C{0.125\columnwidth}|C{0.125\NetTableWidth}|C{0.125\NetTableWidth}|C{0.125\NetTableWidth}|C{0.1\NetTableWidth}|}
    \hhline{~-----~}
    \multicolumn{1}{c|}{} &
    \multicolumn{5}{c|}{\cellcolor{tabhead}Predicted Class} & \multicolumn{1}{c}{}  \\
    \hline
    \rowcolor{tabhead}
    Correct &  Normal & DoS (Hulk) & DoS (Slowloris) & FTP & SSH & Overall \\ \hline
    \cellcolor{tabhead}Normal & 4711 \newline \cellcolor{tabhead}\textbf{(78.52\%)} & 9 \newline \textit{(0.15\%)} & 103 \newline \textit{(1.72\%)} & 148 \newline \textit{(2.47\%)} & 1029 \newline \textit{(17.15\%)} & \multirow{5}{*}{81.28\%}\\ \cline{1-6}
    \cellcolor{tabhead}DoS (Hulk) & 93 \newline \textit{(1.55\%)} & 5745 \newline \cellcolor{tabhead}\textbf{(95.75\%)} & 33 \newline \textit{(0.55\%)} & 43 \newline \textit{(0.72\%)} & 86 \newline \textit{(1.43\%)} & \\ \cline{1-6}
    \cellcolor{tabhead}DoS (Slowloris) & 507 \newline \textit{(8.45\%)} & 0 \newline \textit{(0\%)} & 4668 \newline \cellcolor{tabhead}\textbf{(77.8\%)} & 143 \newline \textit{(2.38\%)} & 682 \newline \textit{(11.37\%)} & \\ \cline{1-6}
    \cellcolor{tabhead}FTP & 643 \newline \textit{(10.72\%)} & 1 \newline \textit{(0.02\%)} & 127 \newline \textit{(2.12\%)} & 4879 \newline \cellcolor{tabhead}\textbf{(81.32\%)} & 350 \newline \textit{(5.83\%)} & \\ \cline{1-6}
    \cellcolor{classcol}SSH & 924 \newline \textit{(15.4\%)} & 34 \newline \textit{(0.57\%)} & 310 \newline \textit{(5.17\%)} & 350 \newline \textit{(5.83\%)} & 4382 \newline \cellcolor{classcol}\textbf{(73.03\%)} & \\ \hline

    \end{tabular}

    \label{tab:cicids-ssh-one-shot}
\end{table}

\begin{table}[H]
    \caption{CICIDS2017 One-Shot Accuracy (SSH not in Training) Using Different $j$ Votes}
    \centering
    \begin{tabular}{|C{0.16\NetTableWidth}|C{0.16\NetTableWidth}|C{0.16\NetTableWidth}|C{0.16\NetTableWidth}|C{0.16\NetTableWidth}|C{0.16\NetTableWidth}|}
        \hline
        \rowcolor{tabhead}
         No Votes & Overall & \multicolumn{2}{c|}{New Class (SSH)} &  \multicolumn{2}{c|}{Normal} \\ \cline{3-6} 
        \rowcolor{tabhead}
        ($j$) & Accuracy&  TPR & FNR & TNR & FPR \\  \hline
      \cellcolor{tabhead} 1 &  72.72\%  &  64.10\% & 16.43\%  & 63.35\% & 36.65\% \\ \hline 
\cellcolor{tabhead} 5 &  81.28\%  &  73.03\%  & 15.40\%  & 78.52\% & 21.48\%\\ \hline 
\cellcolor{tabhead} 10 &  82.56\%  &  77.82\%  & 13.40\%  & 79.95\% & 20.05\%\\ \hline 
\cellcolor{tabhead} 15 &  82.58\%  &  78.43\%  & 13.03\%  & 79.92\% & 20.08\%\\ \hline 
\cellcolor{tabhead} 20 &  82.49\%  &  78.33\%  & 13.18\%  & 79.97\% & 20.03\%\\ \hline 
\cellcolor{tabhead} 25 &  82.43\%  &  78.30\%  & 13.25\%  & 79.78\% & 20.22\%\\ \hline 
\cellcolor{tabhead} 30 &  82.49\% &  78.45\%  & 13.13\%  & 79.97\% & 20.03\%\\ \hline 

    \end{tabular}
    
    \label{tab:cicids-ssh-one-shot-pairs}
    
\end{table}

         
    
    

\begin{table}[h]
    \centering
    \caption{CICIDS2017 One-Shot Confusion Matrix (FTP Not in Training)}
    \begin{tabular}{|C{0.15\NetTableWidth}|C{0.125\NetTableWidth}|C{0.125\NetTableWidth}|C{0.125\columnwidth}|C{0.125\NetTableWidth}|C{0.125\NetTableWidth}|C{0.125\NetTableWidth}|C{0.1\NetTableWidth}|}
    \hhline{~-----~}
    \multicolumn{1}{c|}{} &
    \multicolumn{5}{c|}{\cellcolor{tabhead}Predicted Class} & \multicolumn{1}{|c}{}  \\
    \hline
    \rowcolor{tabhead}
     Correct &  Normal & DoS (Hulk) & DoS (Slowloris) & FTP & SSH & Overall \\ \hline
    \cellcolor{tabhead}Normal & 5231 \newline \cellcolor{tabhead}\textbf{(87.18\%)} & 3 \newline \textit{(0.05\%)} & 152 \newline \textit{(2.53\%)} & 189 \newline \textit{(3.15\%)} & 425 \newline \textit{(7.08\%)} & \multirow{5}{*}{82.5\%}\\ \cline{1-6}
    \cellcolor{tabhead}DoS (Hulk) & 70 \newline \textit{(1.17\%)} & 5755 \newline \cellcolor{tabhead}\textbf{(95.92\%)} & 48 \newline \textit{(0.8\%)} & 15 \newline \textit{(0.25\%)} & 112 \newline \textit{(1.87\%)} & \\ \cline{1-6}
    \cellcolor{tabhead}DoS (Slowloris) & 424 \newline \textit{(7.07\%)} & 1 \newline \textit{(0.02\%)} & 4433 \newline \cellcolor{tabhead}\textbf{(73.88\%)} & 485 \newline \textit{(8.08\%)} & 657 \newline \textit{(10.95\%)} & \\ \cline{1-6}
    \cellcolor{classcol}FTP & 518 \newline \textit{(8.63\%)} & 1 \newline \textit{(0.02\%)} & 659 \newline \textit{(10.98\%)} & 4202 \newline \cellcolor{classcol}\textbf{(70.03\%)} & 620 \newline \textit{(10.33\%)} & \\ \cline{1-6}
    \cellcolor{tabhead}SSH & 546 \newline \textit{(9.1\%)} & 3 \newline \textit{(0.05\%)} & 198 \newline \textit{(3.3\%)} & 124 \newline \textit{(2.07\%)} & 5129 \newline \cellcolor{tabhead}\textbf{(85.48\%)} & \\ \hline

    \end{tabular}

    \label{tab:cicids-ftp-one-shot}
\end{table}

\begin{table}[H]
    \caption{CICIDS2017 One-Shot Accuracy (FTP not in Training) Using Different $j$ Votes}
    \centering
    \begin{tabular}{|C{0.16\NetTableWidth}|C{0.16\NetTableWidth}|C{0.16\NetTableWidth}|C{0.16\NetTableWidth}|C{0.16\NetTableWidth}|C{0.16\NetTableWidth}|}
        \hline
        \rowcolor{tabhead}
         No Votes & Overall & \multicolumn{2}{c|}{New Class (FTP)} &  \multicolumn{2}{c|}{Normal} \\ \cline{3-6} 
        \rowcolor{tabhead}
        ($j$) & Accuracy&  TPR & FNR & TNR & FPR \\  \hline
    \cellcolor{tabhead} 1 &  72.91\%  &  59.65\%  & 8.03\% & 72.83\% & 27.17\% \\ \hline 
    \cellcolor{tabhead} 5 &  82.5\%  &  70.03\% & 8.63\%  & 87.18\% & 12.82\% \\ \hline 
    \cellcolor{tabhead} 10 &  84.57\%  &  72.8\%  & 8.32\%  & 87.70\% & 12.30\% \\ \hline 
    \cellcolor{tabhead} 15 &  85.47\%  &  76.72\%  & 8.12\%  & 87.40\% & 12.60\% \\ \hline 
    \cellcolor{tabhead} 20 &  85.78\%  &  77.58\% & 8.10\%  & 87.23\% & 12.77\% \\ \hline 
    \cellcolor{tabhead} 25 &  85.86\%  &  78.27\%  & 8.10\%  & 86.92\% & 13.08\% \\ \hline 
    \cellcolor{tabhead} 30 &  85.94\% &  78.48\% & 8.00\%  & 86.73\% & 13.27\%\\ \hline 

    \end{tabular}
    
    \label{tab:cicids-ftp-one-shot-pairs}
    
\end{table}

         
    
    

\begin{table}[H]
    \centering
    \caption{CICIDS2017 One-Shot Confusion Matrix (DoS(Hulk) Not in Training)}
    \begin{tabular}{|C{0.15\NetTableWidth}|C{0.125\NetTableWidth}|C{0.125\NetTableWidth}|C{0.125\columnwidth}|C{0.125\NetTableWidth}|C{0.125\NetTableWidth}|C{0.125\NetTableWidth}|C{0.1\NetTableWidth}|}
    \hhline{~-----~}
    \multicolumn{1}{c|}{} &
    \multicolumn{5}{c|}{\cellcolor{tabhead}Predicted Class} & \multicolumn{1}{|c}{}  \\
    \hline
    \rowcolor{tabhead}
    Correct &  Normal & DoS (Hulk) & DoS (Slowloris) & FTP & SSH & Overall \\ \hline
   \cellcolor{tabhead}Normal & 4314 \newline \cellcolor{tabhead}\textbf{(71.9\%)} & 1095 \newline \textit{(18.25\%)} & 174 \newline \textit{(2.9\%)} & 113 \newline \textit{(1.88\%)} & 304 \newline \textit{(5.07\%)} & \multirow{5}{*}{80.81\%}\\ \cline{1-6}
    \cellcolor{classcol}DoS (Hulk) & 78 \newline \textit{(1.3\%)} & 5708 \newline \cellcolor{classcol}\textbf{(95.13\%)} & 60 \newline \textit{(1\%)} & 58 \newline \textit{(0.97\%)} & 96 \newline \textit{(1.6\%)} & \\ \cline{1-6}
    \cellcolor{tabhead}DoS (Slowloris) & 451 \newline \textit{(7.52\%)} & 51 \newline \textit{(0.85\%)} & 4767 \newline \cellcolor{tabhead}\textbf{(79.45\%)} & 111 \newline \textit{(1.85\%)} & 620 \newline \textit{(10.33\%)} & \\ \cline{1-6}
    \cellcolor{tabhead}FTP & 624 \newline \textit{(10.4\%)} & 171 \newline \textit{(2.85\%)} & 138 \newline \textit{(2.3\%)} & 4521 \newline \cellcolor{tabhead}\textbf{(75.35\%)} & 546 \newline \textit{(9.1\%)} & \\ \cline{1-6}
    \cellcolor{tabhead}SSH & 597 \newline \textit{(9.95\%)} & 26 \newline \textit{(0.43\%)} & 245 \newline \textit{(4.08\%)} & 198 \newline \textit{(3.3\%)} & 4934 \newline \cellcolor{tabhead}\textbf{(82.23\%)} & \\   \hline

    \end{tabular}

    \label{tab:cicids-hulk-one-shot}
\end{table}

\begin{table}[H]
    \caption{CICIDS2017 One-Shot Accuracy (DoS (Hulk) not in Training) Using Different $j$ Votes}
    \centering
    \begin{tabular}{|C{0.16\NetTableWidth}|C{0.16\NetTableWidth}|C{0.16\NetTableWidth}|C{0.16\NetTableWidth}|C{0.16\NetTableWidth}|C{0.16\NetTableWidth}|}
        \hline
        \rowcolor{tabhead}
         No Votes & Overall & \multicolumn{2}{c|}{New Class (Hulk)} &  \multicolumn{2}{c|}{Normal} \\ \cline{3-6} 
        \rowcolor{tabhead}
        ($j$) & Accuracy&  TPR & FNR & TNR & FPR \\  \hline
\cellcolor{tabhead} 1 &  72.28\%  &  91.07\% & 4.90\%   & 58.05\% & 41.95\% \\ \hline 
\cellcolor{tabhead} 5 &  80.81\%  &  95.13\%  & 1.30\%  & 71.90\% & 28.10\% \\ \hline 
\cellcolor{tabhead} 10 &  82.59\%  &  95.22\%  & 1.22\%  & 75.58\% & 24.42\%\\ \hline 
\cellcolor{tabhead} 15 &  82.54\%  &  95.23\%  & 1.20\%  & 74.67\% & 25.33\%\\ \hline 
\cellcolor{tabhead} 20 &  82.86\%  &  95.2\% & 1.20\%  & 76.02\% & 23.98\%\\ \hline 
\cellcolor{tabhead} 25 &  82.76\%  &  95.2\%  & 1.15\%  & 75.50\% & 24.50\%\\ \hline 
\cellcolor{tabhead} 30 &  82.93\%  &  95.18\% & 1.22\%  & 76.15\% & 23.85\%\\ \hline 

    \end{tabular}
    
    \label{tab:cicids-hulk-one-shot-pairs}
    
\end{table}

         
    
    

\begin{table}[H]
    \centering
    \caption{CICIDS2017 One-Shot Confusion Matrix (Dos(Slowloris) Not in Training)}
    \begin{tabular}{|C{0.15\NetTableWidth}|C{0.125\NetTableWidth}|C{0.125\NetTableWidth}|C{0.125\columnwidth}|C{0.125\NetTableWidth}|C{0.125\NetTableWidth}|C{0.125\NetTableWidth}|C{0.1\NetTableWidth}|}
    \hhline{~-----~}
    \multicolumn{1}{c|}{} &
    \multicolumn{5}{c|}{\cellcolor{tabhead}Predicted Class} & \multicolumn{1}{|c}{}  \\
    \hline
    \rowcolor{tabhead}
    Correct &  Normal & DoS (Hulk) & DoS (Slowloris) & FTP & SSH & Overall \\ \hline
    \cellcolor{tabhead}Normal & 5307 \newline \cellcolor{tabhead}\textbf{(88.45\%)} & 6 \newline \textit{(0.1\%)} & 459 \newline \textit{(7.65\%)} & 64 \newline \textit{(1.07\%)} & 164 \newline \textit{(2.73\%)} & \multirow{5}{*}{81.07\%}\\ \cline{1-6}
    \cellcolor{tabhead}DoS (Hulk) & 37 \newline \textit{(0.62\%)} & 5794 \newline \cellcolor{tabhead}\textbf{(96.57\%)} & 65 \newline \textit{(1.08\%)} & 53 \newline \textit{(0.88\%)} & 51 \newline \textit{(0.85\%)} & \\ \cline{1-6}
    \cellcolor{classcol}DoS (Slowloris) & 574 \newline \textit{(9.57\%)} & 26 \newline \textit{(0.43\%)} & 4024 \newline \cellcolor{classcol}\textbf{(67.07\%)} & 582 \newline \textit{(9.7\%)} & 794 \newline \textit{(13.23\%)} & \\ \cline{1-6}
    \cellcolor{tabhead}FTP & 482 \newline \textit{(8.03\%)} & 1 \newline \textit{(0.02\%)} & 598 \newline \textit{(9.97\%)} & 4639 \newline \cellcolor{tabhead}\textbf{(77.32\%)} & 280 \newline \textit{(4.67\%)} & \\ \cline{1-6}
    \cellcolor{tabhead}SSH & 446 \newline \textit{(7.43\%)} & 0 \newline \textit{(0\%)} & 817 \newline \textit{(13.62\%)} & 181 \newline \textit{(3.02\%)} & 4556 \newline \cellcolor{tabhead}\textbf{(75.93\%)} & \\ \hline
    \end{tabular}

    \label{tab:cicids-slowloris-one-shot}
\end{table}

\begin{table}[H]
    \caption{CICIDS2017 One-Shot Accuracy (DoS (Slowloris) not in Training) Using Different $j$ Votes}
    \centering
    \begin{tabular}{|C{0.16\NetTableWidth}|C{0.16\NetTableWidth}|C{0.16\NetTableWidth}|C{0.16\NetTableWidth}|C{0.16\NetTableWidth}|C{0.16\NetTableWidth}|}
        \hline
        \rowcolor{tabhead}
         No Votes & Overall & \multicolumn{2}{c|}{New Class (Slowloris)} &  \multicolumn{2}{c|}{Normal} \\ \cline{3-6} 
        \rowcolor{tabhead}
        ($j$) & Accuracy&  TPR & FNR & TNR & FPR \\  \hline
\cellcolor{tabhead} 1 &  72.28\%  &  50.97\%  & 11.50\%  & 72.65\% & 27.35\% \\ \hline 
\cellcolor{tabhead} 5 &  80.81\%  &  67.07\%  & 9.57\%  & 88.45\% & 11.55\%\\ \hline 
\cellcolor{tabhead} 10 &  82.59\%  &  71.38\%  & 7.38\%  & 89.48\% & 10.52\% \\ \hline 
\cellcolor{tabhead} 15 &  82.54\%  &  72.2\%  & 7.18\%  & 89.37\% & 10.63\% \\ \hline 
\cellcolor{tabhead} 20 &  82.86\%  &  72.77\%  & 6.85\%  & 89.67\% & 10.33\% \\ \hline 
\cellcolor{tabhead} 25 &  82.76\%  &  72.93\%  & 6.58\%  & 89.65\% & 10.35\% \\ \hline 
\cellcolor{tabhead} 30 &  82.93\%  &  72.82\%  & 6.68\%  & 89.70\% & 10.30\% \\ \hline 

    \end{tabular}
    
    \label{tab:cicids-slowloris-one-shot-pairs}
    
\end{table}

         
    
    

The CMs of the KDD Cup'99 and NSL-KDD data sets One-Shot, excluding the DoS attack from training are presented in Table~\ref{tab:kdd-dos-one-shot} and Table~\ref{tab:nsl-kdd-dos-one-shot}, respectively; the overall accuracies are 76.67\% and 77.99\%. It is important to note however, that the False Negative rates for the new class (i.e. DoS) are 26.38\% for the KDD Cup'99 and 9.87\% for the NSL-KDD. Additional to the observations arising from the CICIDS2017 evaluation, these results highlight two further elements; (a)~the Siamese Network did not find a high similarity between the new attack and the normal instances; (b) the new attack class TPR in the NSL-KDD results is significantly higher than KDD Cup'99 (78.87\% compared to 40.28\%), because the NSL-KDD is an enhanced version of the KDD Cup'99 (filtered and duplicate instances removed). 
Knowing that the new class is not used in the training phase and the similarity is only calculated from a few instances, a better representation of instances improves performance (i.e. NSL-KDD instances). Results confirm that new labelled instances need to be appropriate representatives. 

\vspace{2mm}
In consideration of completeness, the remaining NSL-KDD and the KDD Cup'99 results - which demonstrate similar performance - are listed as follows; excluding Probe results are listed in Table~\ref{tab:nsl-kdd-probe-one-shot},  Table~\ref{tab:nsl-kdd-probe-one-shot-pairs}, Table~\ref{tab:kdd-probe-one-shot} and Table~\ref{tab:kdd-probe-one-shot-pairs};  Table~\ref{tab:nsl-kdd-u2r-one-shot},  Table~\ref{tab:nsl-kdd-u2r-one-shot-pairs}, Table~\ref{tab:kdd-u2r-one-shot} and Table~\ref{tab:kdd-u2r-one-shot-pairs} present the results when excluding R2L; Finally, excluding U2R are in  Table~\ref{tab:nsl-kdd-r2l-one-shot},  Table~\ref{tab:nsl-kdd-r2l-one-shot-pairs}, Table~\ref{tab:kdd-r2l-one-shot} and Table~\ref{tab:kdd-r2l-one-shot-pairs}.

\begin{table}[htb]
    \centering
    \caption{KDD One-Shot Confusion Matrix (DoS Not in Training)}
    \begin{tabular}{|C{0.15\NetTableWidth}|C{0.14\NetTableWidth}|C{0.14\NetTableWidth}|C{0.14\columnwidth}|C{0.14\NetTableWidth}|C{0.14\NetTableWidth}|C{0.1\NetTableWidth}|}
    \hhline{~-----~}
    \multicolumn{1}{c|}{} &
    \multicolumn{5}{c|}{\cellcolor{tabhead}Predicted Class} & \multicolumn{1}{|c}{}  \\
    \hline
    \rowcolor{tabhead}
    Correct  &  Normal & DoS & Probe & R2L & U2R & Overall \\ \hline
   \cellcolor{tabhead}Normal & 4562 \newline \cellcolor{tabhead}\textbf{(76.03\%)} & 243 \newline \textit{(4.05\%)} & 522 \newline \textit{(8.7\%)} & 579 \newline \textit{(9.65\%)} & 94 \newline \textit{(1.57\%)} & \multirow{5}{*}{76.67\%}\\ \cline{1-6}
    \cellcolor{classcol}DoS & 1583 \newline \textit{(26.38\%)} & 2417 \newline \cellcolor{classcol}\textbf{(40.28\%)} & 1831 \newline \textit{(30.52\%)} & 168 \newline \textit{(2.8\%)} & 1 \newline \textit{(0.02\%)} & \\ \cline{1-6}
    \cellcolor{tabhead}Probe & 159 \newline \textit{(2.65\%)} & 214 \newline \textit{(3.57\%)} & 5367 \newline \cellcolor{tabhead}\textbf{(89.45\%)} & 242 \newline \textit{(4.03\%)} & 18 \newline \textit{(0.3\%)} & \\ \cline{1-6}
    \cellcolor{tabhead}R2L & 56 \newline \textit{(0.93\%)} & 275 \newline \textit{(4.58\%)} & 10 \newline \textit{(0.17\%)} & 5571 \newline \cellcolor{tabhead}\textbf{(92.85\%)} & 88 \newline \textit{(1.47\%)} & \\ \cline{1-6}
    \cellcolor{tabhead}U2R & 17 \newline \textit{(0.28\%)} & 205 \newline \textit{(3.42\%)} & 655 \newline \textit{(10.92\%)} & 40 \newline \textit{(0.67\%)} & 5083 \newline \cellcolor{tabhead}\textbf{(84.72\%)} & \\ \hline
    \end{tabular}
    \label{tab:kdd-dos-one-shot}
\end{table}

\begin{table}[H]
    \caption{KDD One-Shot Accuracy (DoS not in Training) Using Different $j$ Votes}
    \centering
    \begin{tabular}{|C{0.16\NetTableWidth}|C{0.16\NetTableWidth}|C{0.16\NetTableWidth}|C{0.16\NetTableWidth}|C{0.16\NetTableWidth}|C{0.16\NetTableWidth}|}
        \hline
        \rowcolor{tabhead}
         No Votes & Overall & \multicolumn{2}{c|}{New Class (DoS)} &  \multicolumn{2}{c|}{Normal} \\ \cline{3-6} 
        \rowcolor{tabhead}
        ($j$) & Accuracy&  TPR & FNR & TNR & FPR \\  \hline
\cellcolor{tabhead} 1 &  66.89\%  &  41.67\%  & 22.50\%   & 66.35\% & 33.65\% \\ \hline 
\cellcolor{tabhead} 5 &  76.67\%  &  40.28\%  & 26.38\%  & 76.03\% & 23.97\% \\ \hline 
\cellcolor{tabhead} 10 &  77.57\%  &  40.07\%  & 27.25\%  & 76.10\% & 23.90\% \\ \hline 
\cellcolor{tabhead} 15 &  77.67\%  &  39.9\%  & 27.32\%  & 76.02\% & 23.98\% \\ \hline 
\cellcolor{tabhead} 20 &  77.68\%  &  39.93\%  & 27.38\%  & 76.02\% & 23.98\% \\ \hline 
\cellcolor{tabhead} 25 &  77.68\%  &  39.87\%  & 27.40\%  & 76.07\% & 23.93\% \\ \hline 
\cellcolor{tabhead} 30 &  77.68\% &  39.88\%  & 27.40\%  & 76.03\% & 23.97\% \\ \hline 

    \end{tabular}
    
    \label{tab:kdd-dos-one-shot-pairs}
    
\end{table}


         
    
\begin{table}[!htb]
    \centering
    \caption{NSL-KDD One-Shot Confusion Matrix (DoS Not in Training)}
    \begin{tabular}{|C{0.15\NetTableWidth}|C{0.14\NetTableWidth}|C{0.14\NetTableWidth}|C{0.14\columnwidth}|C{0.14\NetTableWidth}|C{0.14\NetTableWidth}|C{0.1\NetTableWidth}|}
    \hhline{~-----~}
    \multicolumn{1}{c|}{} &
    \multicolumn{5}{c|}{\cellcolor{tabhead}Predicted Class} & \multicolumn{1}{|c}{}  \\
    \hline
    \rowcolor{tabhead}
    Correct  &  Normal & DoS & Probe & R2L & U2R & Overall \\ \hline
    \cellcolor{tabhead}Normal & 5593 \newline \cellcolor{tabhead}\textbf{(93.22\%)} & 61 \newline \textit{(1.02\%)} & 136 \newline \textit{(2.27\%)} & 122 \newline \textit{(2.03\%)} & 88 \newline \textit{(1.47\%)} & \multirow{5}{*}{77.99\%}\\ \cline{1-6}
    \cellcolor{classcol}DoS & 592 \newline \textit{(9.87\%)} & 4732 \newline \cellcolor{classcol}\textbf{(78.87\%)} & 653 \newline \textit{(10.88\%)} & 12 \newline \textit{(0.2\%)} & 11 \newline \textit{(0.18\%)} & \\ \cline{1-6}
    \cellcolor{tabhead}Probe & 67 \newline \textit{(1.12\%)} & 3305 \newline \textit{(55.08\%)} & 2595 \newline \cellcolor{tabhead}\textbf{(43.25\%)} & 19 \newline \textit{(0.32\%)} & 14 \newline \textit{(0.23\%)} & \\ \cline{1-6}
    \cellcolor{tabhead}R2L & 212 \newline \textit{(3.53\%)} & 7 \newline \textit{(0.12\%)} & 27 \newline \textit{(0.45\%)} & 5692 \newline \cellcolor{tabhead}\textbf{(94.87\%)} & 62 \newline \textit{(1.03\%)} & \\ \cline{1-6}
    \cellcolor{tabhead}U2R & 486 \newline \textit{(8.1\%)} & 6 \newline \textit{(0.1\%)} & 31 \newline \textit{(0.52\%)} & 693 \newline \textit{(11.55\%)} & 4784 \newline \cellcolor{tabhead}\textbf{(79.73\%)} & \\ \hline
    \end{tabular}
    \label{tab:nsl-kdd-dos-one-shot}
\end{table}

\begin{table}[H]
    \caption{NSL-KDD One-Shot Accuracy (DoS not in Training) Using Different $j$ Votes}
    \centering
    \begin{tabular}{|C{0.16\NetTableWidth}|C{0.16\NetTableWidth}|C{0.16\NetTableWidth}|C{0.16\NetTableWidth}|C{0.16\NetTableWidth}|C{0.16\NetTableWidth}|}
        \hline
        \rowcolor{tabhead}
         No Votes & Overall & \multicolumn{2}{c|}{New Class (DoS)} &  \multicolumn{2}{c|}{Normal} \\ \cline{3-6} 
        \rowcolor{tabhead}
        ($j$) & Accuracy&  TPR & FNR & TNR & FPR \\  \hline
\cellcolor{tabhead} 1 &  72.75\%  &  67.35\% & 9.05\%  & 84.87\% & 15.13\%\\ \hline 
\cellcolor{tabhead} 5 &  77.99\%  &  78.87\%  & 9.87\%  & 93.22\% & 6.78\% \\ \hline 
\cellcolor{tabhead} 10 &  77.7\%  &  84.62\%  & 9.87\%  & 93.35\% & 6.65\% \\ \hline 
\cellcolor{tabhead} 15 &  79.05\%  &  83.78\% & 9.87\%  & 93.32\% & 6.68\% \\ \hline 
\cellcolor{tabhead} 20 &  78.63\%  &  85.25\% & 9.87\% & 93.37\% & 6.63\% \\ \hline 
\cellcolor{tabhead} 25 &  79.49\%  &  84.62\%  & 9.87\% & 93.35\% & 6.65\% \\ \hline 
\cellcolor{tabhead} 30 &  79.12\%  &  85.37\%  & 9.87\% & 93.35\% & 6.65\% \\ \hline 

    \end{tabular}
    
    \label{tab:nsl-kdd-dos-one-shot-pairs}
    
\end{table}

    

\begin{table}[H]
\centering
    \caption{NSL-KDD One-Shot Confusion Matrix (Probe Not in Training)}
    \begin{tabular}{|C{0.15\NetTableWidth}|C{0.14\NetTableWidth}|C{0.14\NetTableWidth}|C{0.14\columnwidth}|C{0.14\NetTableWidth}|C{0.14\NetTableWidth}|C{0.1\NetTableWidth}|}
    \hhline{~-----~}
    \multicolumn{1}{c|}{} &
    \multicolumn{5}{c|}{\cellcolor{tabhead}Predicted Class} & \multicolumn{1}{|c}{}  \\
    \hline
    \rowcolor{tabhead}
    Correct  &  Normal & DoS & Probe & R2L & U2R & Overall \\ \hline
     \cellcolor{tabhead}Normal & 5389 \newline \cellcolor{tabhead}\textbf{(89.82\%)} & 89 \newline \textit{(1.48\%)} & 195 \newline \textit{(3.25\%)} & 245 \newline \textit{(4.08\%)} & 82 \newline \textit{(1.37\%)} & \multirow{5}{*}{75.31\%}\\ \cline{1-6}
    \cellcolor{tabhead}DoS & 37 \newline \textit{(0.62\%)} & 5842 \newline \cellcolor{tabhead}\textbf{(97.37\%)} & 95 \newline \textit{(1.58\%)} & 21 \newline \textit{(0.35\%)} & 5 \newline \textit{(0.08\%)} & \\ \cline{1-6}
    \cellcolor{classcol}Probe & 1697 \newline \textit{(28.28\%)} & 2571 \newline \textit{(42.85\%)} & 565 \newline \cellcolor{classcol}\textbf{(9.42\%)} & 948 \newline \textit{(15.8\%)} & 219 \newline \textit{(3.65\%)} & \\ \cline{1-6}
    \cellcolor{tabhead}R2L & 54 \newline \textit{(0.9\%)} & 0 \newline \textit{(0\%)} & 55 \newline \textit{(0.92\%)} & 5800 \newline \cellcolor{tabhead}\textbf{(96.67\%)} & 91 \newline \textit{(1.52\%)} & \\ \cline{1-6}
    \cellcolor{tabhead}U2R & 263 \newline \textit{(4.38\%)} & 0 \newline \textit{(0\%)} & 21 \newline \textit{(0.35\%)} & 720 \newline \textit{(12\%)} & 4996 \newline \cellcolor{tabhead}\textbf{(83.27\%)} & \\ \hline
    \end{tabular}
    \label{tab:nsl-kdd-probe-one-shot}
\end{table}

\begin{table}[H]
    \caption{NSL-KDD One-Shot Accuracy (Probe not in Training) Using Different $j$ Votes}
    \centering
    \begin{tabular}{|C{0.16\NetTableWidth}|C{0.16\NetTableWidth}|C{0.16\NetTableWidth}|C{0.16\NetTableWidth}|C{0.16\NetTableWidth}|C{0.16\NetTableWidth}|}
        \hline
        \rowcolor{tabhead}
         No Votes & Overall & \multicolumn{2}{c|}{New Class (Probe)} &  \multicolumn{2}{c|}{Normal} \\ \cline{3-6} 
        \rowcolor{tabhead}
        ($j$) & Accuracy&  TPR & FNR & TNR & FPR \\  \hline
\cellcolor{tabhead} 1 &  70.62\%  &  18.80\% & 24.78\%  & 77.53\% & 22.47\% \\ \hline 
\cellcolor{tabhead} 5 &  75.31\%  &  9.42\%  & 28.28\%  & 89.82\% & 10.18\%\\ \hline 
\cellcolor{tabhead} 10 &  75.2\%  &  4.83\%  & 28.82\% & 91.08\% & 8.92\%  \\ \hline 
\cellcolor{tabhead} 15 &  75.12\%  &  4.05\%  & 29.08\%  & 91.18\% & 8.82\%
 \\ \hline 
\cellcolor{tabhead} 20 &  75.11\%  &  3.47\%  & 29.20\%  & 91.45\% & 8.55\% \\ \hline 
\cellcolor{tabhead} 25 &  75\%  &  3.02\%  & 29.55\%  & 91.35\% & 8.65\%
\\ \hline 
\cellcolor{tabhead} 30 &  74.94\%  &  2.68\%  & 29.68\%  & 91.33\% & 8.67\%\\ \hline 

    \end{tabular}
    
    \label{tab:nsl-kdd-probe-one-shot-pairs}
    
\end{table}
    
\begin{table}[H]
    \centering
    \caption{KDD One-Shot Confusion Matrix (Probe Not in Training)}
    \begin{tabular}{|C{0.15\NetTableWidth}|C{0.14\NetTableWidth}|C{0.14\NetTableWidth}|C{0.14\columnwidth}|C{0.14\NetTableWidth}|C{0.14\NetTableWidth}|C{0.1\NetTableWidth}|}
    \hhline{~-----~}
    \multicolumn{1}{c|}{} &
    \multicolumn{5}{c|}{\cellcolor{tabhead}Predicted Class} & \multicolumn{1}{|c}{}  \\
    \hline
    \rowcolor{tabhead}
    Correct  &  Normal & DoS & Probe & R2L & U2R & Overall \\ \hline
     \cellcolor{tabhead}Normal & 4515 \newline \cellcolor{tabhead}\textbf{(75.25\%)} & 16 \newline \textit{(0.27\%)} & 383 \newline \textit{(6.38\%)} & 1016 \newline \textit{(16.93\%)} & 70 \newline \textit{(1.17\%)} & \multirow{5}{*}{72.23\%}\\ \cline{1-6}
    \cellcolor{tabhead}DoS & 18 \newline \textit{(0.3\%)} & 5896 \newline \cellcolor{tabhead}\textbf{(98.27\%)} & 81 \newline \textit{(1.35\%)} & 4 \newline \textit{(0.07\%)} & 1 \newline \textit{(0.02\%)} & \\ \cline{1-6}
    \cellcolor{classcol}Probe & 719 \newline \textit{(11.98\%)} & 3707 \newline \textit{(61.78\%)} & 612 \newline \cellcolor{classcol}\textbf{(10.2\%)} & 941 \newline \textit{(15.68\%)} & 21 \newline \textit{(0.35\%)} & \\ \cline{1-6}
    \cellcolor{tabhead}R2L & 26 \newline \textit{(0.43\%)} & 0 \newline \textit{(0\%)} & 16 \newline \textit{(0.27\%)} & 5946 \newline \cellcolor{tabhead}\textbf{(99.1\%)} & 12 \newline \textit{(0.2\%)} & \\ \cline{1-6}
    \cellcolor{tabhead}U2R & 55 \newline \textit{(0.92\%)} & 37 \newline \textit{(0.62\%)} & 264 \newline \textit{(4.4\%)} & 943 \newline \textit{(15.72\%)} & 4701 \newline \cellcolor{tabhead}\textbf{(78.35\%)} & \\ \hline
    \end{tabular}
    \label{tab:kdd-probe-one-shot}
\end{table}

\begin{table}[H]
    \caption{KDD One-Shot Accuracy (Probe not in Training) Using Different $j$ Votes}
    \centering
    \begin{tabular}{|C{0.16\NetTableWidth}|C{0.16\NetTableWidth}|C{0.16\NetTableWidth}|C{0.16\NetTableWidth}|C{0.16\NetTableWidth}|C{0.16\NetTableWidth}|}
        \hline
        \rowcolor{tabhead}
         No Votes & Overall & \multicolumn{2}{c|}{New Class (Probe)} &  \multicolumn{2}{c|}{Normal} \\ \cline{3-6} 
        \rowcolor{tabhead}
        ($j$) & Accuracy&  TPR & FNR & TNR & FPR \\  \hline
\cellcolor{tabhead} 1 &  66.72\%  &  15.72\% & 11.77\%  & 65.72\% & 34.28\% \\ \hline 
\cellcolor{tabhead} 5 &  72.23\%  &  10.2\%  & 11.98\%  & 75.25\% & 24.75\% \\ \hline 
\cellcolor{tabhead} 10 &  72.59\%  &  5.9\%  & 13.30\%  & 78.65\% & 21.35\% \\ \hline 
\cellcolor{tabhead} 15 &  72.35\%  &  4.82\%  & 13.08\%  & 78.57\% & 21.43\% \\ \hline 
\cellcolor{tabhead} 20 &  72.26\%  &  3.58\%  & 13.50\%  & 79.20\% & 20.80\% \\ \hline 
\cellcolor{tabhead} 25 &  72.17\%  &  3.05\%  & 13.55\%  & 79.23\% & 20.77\% \\ \hline 
\cellcolor{tabhead} 30 &  72.07\% &  2.17\% & 13.98\%  & 79.62\% & 20.38\% \\ \hline 

    \end{tabular}
    
    \label{tab:kdd-probe-one-shot-pairs}
    
\end{table}


         
    

\begin{table}[H]
    \centering
    \caption{NSL- KDD One-Shot Confusion Matrix (R2L Not in Training)}
    \begin{tabular}{|C{0.15\NetTableWidth}|C{0.14\NetTableWidth}|C{0.14\NetTableWidth}|C{0.14\columnwidth}|C{0.14\NetTableWidth}|C{0.14\NetTableWidth}|C{0.1\NetTableWidth}|}
    \hhline{~-----~}
    \multicolumn{1}{c|}{} &
    \multicolumn{5}{c|}{\cellcolor{tabhead}Predicted Class} & \multicolumn{1}{|c}{}  \\
    \hline
    \rowcolor{tabhead}
    Correct  &  Normal & DoS & Probe & R2L & U2R & Overall \\ \hline
    \cellcolor{tabhead}Normal & 5199 \newline \cellcolor{tabhead}\textbf{(86.65\%)} & 24 \newline \textit{(0.4\%)} & 148 \newline \textit{(2.47\%)} & 530 \newline \textit{(8.83\%)} & 99 \newline \textit{(1.65\%)} & \multirow{5}{*}{80.16\%}\\ \cline{1-6}
    \cellcolor{tabhead}DoS & 15 \newline \textit{(0.25\%)} & 5799 \newline \cellcolor{tabhead}\textbf{(96.65\%)} & 36 \newline \textit{(0.6\%)} & 26 \newline \textit{(0.43\%)} & 124 \newline \textit{(2.07\%)} & \\ \cline{1-6}
    \cellcolor{tabhead}Probe & 90 \newline \textit{(1.5\%)} & 242 \newline \textit{(4.03\%)} & 5416 \newline \cellcolor{tabhead}\textbf{(90.27\%)} & 236 \newline \textit{(3.93\%)} & 16 \newline \textit{(0.27\%)} & \\ \cline{1-6}
    \cellcolor{classcol}R2L & 2526 \newline \textit{(42.1\%)} & 1 \newline \textit{(0.02\%)} & 142 \newline \textit{(2.37\%)} & 2759 \newline \cellcolor{classcol}\textbf{(45.98\%)} & 572 \newline \textit{(9.53\%)} & \\ \cline{1-6}
    \cellcolor{tabhead}U2R & 852 \newline \textit{(14.2\%)} & 3 \newline \textit{(0.05\%)} & 0 \newline \textit{(0\%)} & 270 \newline \textit{(4.5\%)} & 4875 \newline \cellcolor{tabhead}\textbf{(81.25\%)} & \\ \hline
    \end{tabular}
    \label{tab:nsl-kdd-r2l-one-shot}
\end{table}

\begin{table}[H]
    \caption{NSL-KDD One-Shot Accuracy (R2L not in Training) Using Different $j$ Votes}
    \centering
    \begin{tabular}{|C{0.16\NetTableWidth}|C{0.16\NetTableWidth}|C{0.16\NetTableWidth}|C{0.16\NetTableWidth}|C{0.16\NetTableWidth}|C{0.16\NetTableWidth}|}
        \hline
        \rowcolor{tabhead}
         No Votes & Overall & \multicolumn{2}{c|}{New Class (R2L)} &  \multicolumn{2}{c|}{Normal} \\ \cline{3-6} 
        \rowcolor{tabhead}
        ($j$) & Accuracy&  TPR & FNR & TNR & FPR \\  \hline
\cellcolor{tabhead} 1 &  74.5\%  &  46.05\% & 38.13\%  & 74.73\% & 25.27\% \\ \hline 
\cellcolor{tabhead} 5 &  80.16\%  &  45.98\%  & 42.10\%  & 86.65\% & 13.35\% \\ \hline 
\cellcolor{tabhead} 10 &  80.79\%  &  46.82\%  & 41.58\%  & 88.07\% & 11.93\%\\ \hline 
\cellcolor{tabhead} 15 &  81.09\%  &  49.02\%  & 39.88\%  & 87.72\% & 12.28\%\\ \hline 
\cellcolor{tabhead} 20 &  81\%  &  48.62\%  & 40.38\%  & 87.90\% & 12.10\% \\ \hline 
\cellcolor{tabhead} 25 &  80.95\%  &  48.37\%  & 40.63\%  & 87.88\% & 12.12\% \\ \hline 
\cellcolor{tabhead} 30 &  80.91\%  &  48.2\%  & 40.93\%  & 87.93\% & 12.07\%\\ \hline 

    \end{tabular}
    
    \label{tab:nsl-kdd-r2l-one-shot-pairs}
    
\end{table}

    

\begin{table}[H]
    \centering
    \caption{KDD One-Shot Confusion Matrix (R2L Not in Training)}
    \begin{tabular}{|C{0.15\NetTableWidth}|C{0.14\NetTableWidth}|C{0.14\NetTableWidth}|C{0.14\columnwidth}|C{0.14\NetTableWidth}|C{0.14\NetTableWidth}|C{0.1\NetTableWidth}|}
    \hhline{~-----~}
    \multicolumn{1}{c|}{} &
    \multicolumn{5}{c|}{\cellcolor{tabhead}Predicted Class} & \multicolumn{1}{|c}{}  \\
    \hline
    \rowcolor{tabhead}
    Correct  &  Normal & DoS & Probe & R2L & U2R & Overall \\ \hline
    \cellcolor{tabhead}Normal & 4288 \newline \cellcolor{tabhead}\textbf{(71.47\%)} & 1 \newline \textit{(0.02\%)} & 400 \newline \textit{(6.67\%)} & 730 \newline \textit{(12.17\%)} & 581 \newline \textit{(9.68\%)} & \multirow{5}{*}{74.2\%}\\ \cline{1-6}
    \cellcolor{tabhead}DoS & 10 \newline \textit{(0.17\%)} & 5909 \newline \cellcolor{tabhead}\textbf{(98.48\%)} & 72 \newline \textit{(1.2\%)} & 9 \newline \textit{(0.15\%)} & 0 \newline \textit{(0\%)} & \\ \cline{1-6}
    \cellcolor{tabhead}Probe & 90 \newline \textit{(1.5\%)} & 160 \newline \textit{(2.67\%)} & 5338 \newline \cellcolor{tabhead}\textbf{(88.97\%)} & 165 \newline \textit{(2.75\%)} & 247 \newline \textit{(4.12\%)} & \\ \cline{1-6}
    \cellcolor{classcol}R2L & 1702 \newline \textit{(28.37\%)} & 2 \newline \textit{(0.03\%)} & 1344 \newline \textit{(22.4\%)} & 2148 \newline \cellcolor{classcol}\textbf{(35.8\%)} & 804 \newline \textit{(13.4\%)} & \\ \cline{1-6}
    \cellcolor{tabhead}U2R & 527 \newline \textit{(8.78\%)} & 1 \newline \textit{(0.02\%)} & 682 \newline \textit{(11.37\%)} & 213 \newline \textit{(3.55\%)} & 4577 \newline \cellcolor{tabhead}\textbf{(76.28\%)} & \\ \hline
    \end{tabular}
    \label{tab:kdd-r2l-one-shot}
\end{table}

\begin{table}[H]
    \caption{KDD One-Shot Accuracy (R2L not in Training) Using Different $j$ Votes}
    \centering
    \begin{tabular}{|C{0.16\NetTableWidth}|C{0.16\NetTableWidth}|C{0.16\NetTableWidth}|C{0.16\NetTableWidth}|C{0.16\NetTableWidth}|C{0.16\NetTableWidth}|}
        \hline
        \rowcolor{tabhead}
         No Votes & Overall & \multicolumn{2}{c|}{New Class (R2L)} &  \multicolumn{2}{c|}{Normal} \\ \cline{3-6} 
        \rowcolor{tabhead}
        ($j$) & Accuracy&  TPR & FNR & TNR & FPR \\  \hline
\cellcolor{tabhead} 1 &  67.75\%  &  38.48\% & 25.95\%   & 59.65\% & 40.35\% \\ \hline 
\cellcolor{tabhead} 5 &  74.2\%  &  35.8\% & 28.37\%  & 71.47\% & 28.53\%\\ \hline 
\cellcolor{tabhead} 10 &  77.27\%  &  42.22\% & 23.85\%  & 74.38\% & 25.62\%\\ \hline 
\cellcolor{tabhead} 15 &  78.34\%  &  46.65\%  & 22.05\% & 74.50\% & 25.50\% \\ \hline 
\cellcolor{tabhead} 20 &  78.94\%  &  49.18\%  & 21.45\%  & 74.62\% & 25.38\%\\ \hline 
\cellcolor{tabhead} 25 &  79.44\%  &  51.32\%  & 20.72\%  & 74.65\% & 25.35\% \\ \hline 
\cellcolor{tabhead} 30 &  79.87\%  &  53.35\%  & 20.65\%  & 74.55\% & 25.45\%\\ \hline 

    \end{tabular}
    
    \label{tab:kdd-r2l-one-shot-pairs}
    
\end{table}

         
    
\begin{table}[H]
    \centering
    \caption{NSL-KDD One-Shot Confusion Matrix (U2R Not in Training)}
    \begin{tabular}{|C{0.15\NetTableWidth}|C{0.14\NetTableWidth}|C{0.14\NetTableWidth}|C{0.14\columnwidth}|C{0.14\NetTableWidth}|C{0.14\NetTableWidth}|C{0.1\NetTableWidth}|}
    \hhline{~-----~}
    \multicolumn{1}{c|}{} &
    \multicolumn{5}{c|}{\cellcolor{tabhead}Predicted Class} & \multicolumn{1}{|c}{}  \\
    \hline
    \rowcolor{tabhead}
    Correct  &  Normal & DoS & Probe & R2L & U2R & Overall \\ \hline
    \cellcolor{tabhead}Normal & 4530 \newline \cellcolor{tabhead}\textbf{(75.5\%)} & 127 \newline \textit{(2.12\%)} & 76 \newline \textit{(1.27\%)} & 237 \newline \textit{(3.95\%)} & 1030 \newline \textit{(17.17\%)} & \multirow{5}{*}{77.04\%}\\ \cline{1-6}
    \cellcolor{tabhead}DoS & 120 \newline \textit{(2\%)} & 5771 \newline \cellcolor{tabhead}\textbf{(96.18\%)} & 49 \newline \textit{(0.82\%)} & 16 \newline \textit{(0.27\%)} & 44 \newline \textit{(0.73\%)} & \\ \cline{1-6}
    \cellcolor{tabhead}Probe & 43 \newline \textit{(0.72\%)} & 304 \newline \textit{(5.07\%)} & 5574 \newline \cellcolor{tabhead}\textbf{(92.9\%)} & 69 \newline \textit{(1.15\%)} & 10 \newline \textit{(0.17\%)} & \\ \cline{1-6}
    \cellcolor{tabhead}R2L & 403 \newline \textit{(6.72\%)} & 1 \newline \textit{(0.02\%)} & 27 \newline \textit{(0.45\%)} & 5238 \newline \cellcolor{tabhead}\textbf{(87.3\%)} & 331 \newline \textit{(5.52\%)} & \\ \cline{1-6}
    \cellcolor{classcol}U2R & 2191 \newline \textit{(36.52\%)} & 0 \newline \textit{(0\%)} & 221 \newline \textit{(3.68\%)} & 1589 \newline \textit{(26.48\%)} & 1999 \newline \cellcolor{classcol}\textbf{(33.32\%)} & \\ \hline
    \end{tabular}
    \label{tab:nsl-kdd-u2r-one-shot}
\end{table}

\begin{table}[H]
    \caption{NSL-KDD One-Shot Accuracy (U2R not in Training) Using Different $j$ Votes}
    \centering
    \begin{tabular}{|C{0.16\NetTableWidth}|C{0.16\NetTableWidth}|C{0.16\NetTableWidth}|C{0.16\NetTableWidth}|C{0.16\NetTableWidth}|C{0.16\NetTableWidth}|}
        \hline
        \rowcolor{tabhead}
         No Votes & Overall & \multicolumn{2}{c|}{New Class (U2R)} &  \multicolumn{2}{c|}{Normal} \\ \cline{3-6} 
        \rowcolor{tabhead}
        ($j$) & Accuracy&  TPR & FNR & TNR & FPR \\  \hline
\cellcolor{tabhead} 1 &  72.42\%  &  34.37\% & 35.55\%  & 66.58\% & 33.42\% \\ \hline 
\cellcolor{tabhead} 5 &  77.04\%  &  33.32\% & 36.52\%  & 75.50\% & 24.50\% \\ \hline 
\cellcolor{tabhead} 10 &  77.08\%  &  30.42\%  & 36.95\%  & 77.85\% & 22.15\%\\ \hline 
\cellcolor{tabhead} 15 &  77.19\%  &  30.2\%  & 36.70\%  & 78.22\% & 21.78\%\\ \hline 
\cellcolor{tabhead} 20 &  77.12\%  &  29.37\%  & 36.67\%  & 78.52\% & 21.48\% \\ \hline 
\cellcolor{tabhead} 25 &  77.14\%  &  28.85\%  & 36.72\%  & 78.87\% & 21.13\% \\ \hline 
\cellcolor{tabhead} 30 &  77.12\%  &  28.3\% & 37.10\%  & 79.25\% & 20.75\% \\ \hline 

    \end{tabular}
    
    \label{tab:nsl-kdd-u2r-one-shot-pairs}
    
\end{table}

    
\begin{table}[H]
    \centering
    \caption{KDD One-Shot Confusion Matrix (U2R Not in Training)}
    \begin{tabular}{|C{0.15\NetTableWidth}|C{0.14\NetTableWidth}|C{0.14\NetTableWidth}|C{0.14\columnwidth}|C{0.14\NetTableWidth}|C{0.14\NetTableWidth}|C{0.1\NetTableWidth}|}
    \hhline{~-----~}
    \multicolumn{1}{c|}{} &
    \multicolumn{5}{c|}{\cellcolor{tabhead}Predicted Class} & \multicolumn{1}{|c}{}  \\
    \hline
    \rowcolor{tabhead}
    Correct  &  Normal & DoS & Probe & R2L & U2R & Overall \\ \hline
    \cellcolor{tabhead}Normal & 4146 \newline \cellcolor{tabhead}\textbf{(69.1\%)} & 5 \newline \textit{(0.08\%)} & 440 \newline \textit{(7.33\%)} & 796 \newline \textit{(13.27\%)} & 613 \newline \textit{(10.22\%)} & \multirow{5}{*}{75.72\%}\\ \cline{1-6}
    \cellcolor{tabhead}DoS & 7 \newline \textit{(0.12\%)} & 5921 \newline \cellcolor{tabhead}\textbf{(98.68\%)} & 59 \newline \textit{(0.98\%)} & 6 \newline \textit{(0.1\%)} & 7 \newline \textit{(0.12\%)} & \\ \cline{1-6}
    \cellcolor{tabhead}Probe & 53 \newline \textit{(0.88\%)} & 384 \newline \textit{(6.4\%)} & 5449 \newline \cellcolor{tabhead}\textbf{(90.82\%)} & 59 \newline \textit{(0.98\%)} & 55 \newline \textit{(0.92\%)} & \\ \cline{1-6}
    \cellcolor{tabhead}R2L & 35 \newline \textit{(0.58\%)} & 0 \newline \textit{(0\%)} & 13 \newline \textit{(0.22\%)} & 5849 \newline \cellcolor{tabhead}\textbf{(97.48\%)} & 103 \newline \textit{(1.72\%)} & \\ \cline{1-6}
    \cellcolor{classcol}U2R & 958 \newline \textit{(15.97\%)} & 1 \newline \textit{(0.02\%)} & 669 \newline \textit{(11.15\%)} & 3022 \newline \textit{(50.37\%)} & 1350 \newline \cellcolor{classcol}\textbf{(22.5\%)} & \\ \hline
    \end{tabular}
   \label{tab:kdd-u2r-one-shot}
\end{table}

\begin{table}[H]
    \caption{KDD One-Shot Accuracy (U2R not in Training) Using Different $j$ Votes}
    \centering
     \begin{tabular}{|C{0.16\NetTableWidth}|C{0.16\NetTableWidth}|C{0.16\NetTableWidth}|C{0.16\NetTableWidth}|C{0.16\NetTableWidth}|C{0.16\NetTableWidth}|}
        \hline
        \rowcolor{tabhead}
         No Votes & Overall & \multicolumn{2}{c|}{New Class (U2R)} &  \multicolumn{2}{c|}{Normal} \\ \cline{3-6} 
        \rowcolor{tabhead}
        ($j$) & Accuracy&  TPR & FNR & TNR & FPR \\  \hline
\cellcolor{tabhead} 1 &  70.69\%  &  21.40\% & 17.28\%  & 59.27\% & 40.73\% \\ \hline 
\cellcolor{tabhead} 5 &  75.72\%  &  22.5\% & 15.97\%  & 69.10\% & 30.90\%\\ \hline 
\cellcolor{tabhead} 10 &  76.26\%  &  21.82\% & 17.17\%   & 72.18\% & 27.82\%\\ \hline 
\cellcolor{tabhead} 15 &  76.33\%  &  21.83\%  & 17.15\%  & 72.52\% & 27.48\%\\ \hline 
\cellcolor{tabhead} 20 &  76.31\%  &  21.48\%  & 17.52\%  & 72.72\% & 27.28\%\\ \hline 
\cellcolor{tabhead} 25 &  76.34\%  &  21.45\%  & 17.55\%  & 72.77\% & 27.23\%\\ \hline 
\cellcolor{tabhead} 30 &  76.33\%  &  21.27\% & 17.73\%  & 72.90\% & 27.10\%\\ \hline 

    \end{tabular}
    
    \label{tab:kdd-u2r-one-shot-pairs}
    
\end{table}

\section{Conclusion and Future Work}
\label{sec:conclusion}
The paper presents an Intrusion Detection Siamese Network framework capable of classifying new cyber-attacks based on a limited number of labelled instances (One-Shot). The evaluation of the model was performed on three different data sets; CICIDS2017, KDD Cup'99 and the NSL-KDD, an enhancement of the KDD Cup'99.

Results of the evaluation re-confirm that particular consideration must be given on creating the training set, ensuring an equal number of training pairs for every class combination. The core requirement, in turn, presents a challenge of an exploding number of combinations between all instances. Thus, distinct pairs are chosen to create large batches in the region of 30,000 pairs to mitigate the growth. During evaluation, similarity comparison using a single point for each class resulted in noisy predictions due to randomness obviated through the selection of multiple ($j$) random instances from each class and aggregation using majority voting.

Results demonstrate the ability of the proposed architecture to classify cyber-attacks based on learning from similarity. Moreover, the results highlighted the need for representative instances for the new attack class. Furthermore, evidence is provided to confirm the ability of One-Shot learning methodologies to adapt to new cyber-attacks without retraining when only a few instances are available for a new attack. An overall accuracy of between 80\% - 85\% for the CICIDS2017 data set was evaluated, demonstrating acceptable accuracy in detecting previously unseen attacks. The overall accuracy reached above 75\% for the KDD Cup'99 and NSL-KDD data sets. Further and also important to the application is that the overall accuracy was achieved at a low FNR for the new attack classes.  

\bibliographystyle{IEEEtran}
\bibliography{bibliography}

%

\newpage
\begin{IEEEbiography}[{\includegraphics[width=1in,height=1.25in,clip,keepaspectratio]{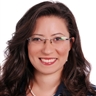}}]{Hanan Hindy}(S$'$18) is a third year PhD student at the Division of Cyber-Security at Abertay University, Dundee, Scotland. Hanan received her bachelor degree with honours~(2012) and masters~(2016) degrees in Computer Science from the Faculty of Computer and Information Sciences at Ain Shams University, Cairo, Egypt. \\
Her research interests include Machine Learning and Cyber-Security. Currently, she is working on utilising deep learning for IDS.
\end{IEEEbiography}

\begin{IEEEbiography}[{\includegraphics[width=1in,height=1.25in,clip,keepaspectratio]{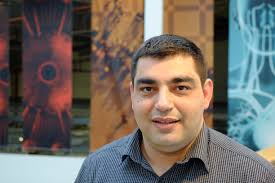}}]{Christos Tachtatzis}is a Senior Lecturer Chancellor’s Fellow in Sensor Systems and Asset Management, at the University of Strathclyde. He holds a BEng (Hons) in Communication Systems Engineering from University of Portsmouth in 2001, an MSc in Communications, Control and Digital Signal Processing (2002) and a PhD in Electronic and Electrical Engineering (2008), both from Strathclyde University. Christos has 12 years experience, in Sensor Systems ranging from electronic devices, networking, communications and signal processing. His current research interests lie in extracting actionable information from data using machine learning and artificial intelligence.
\end{IEEEbiography}

\begin{IEEEbiography}[{\includegraphics[width=1in,height=1.25in,clip,keepaspectratio]{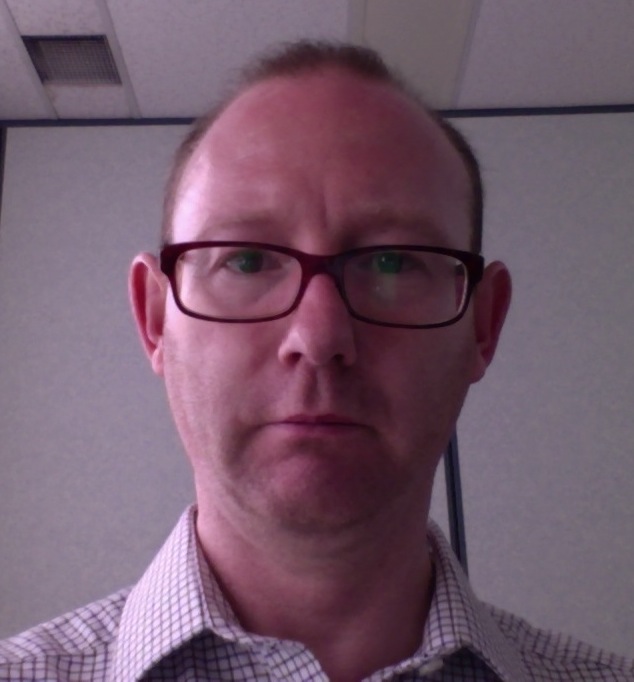}}]{Robert C. Atkinson}
(M$'$03 -- S$'$98 -- SM$'$07) received the B.Eng. (Hons.) degree in electronic and electrical engineering; the M.Sc. degree in communications, control, and digital signal processing; and the Ph.D. degree in mobile communications systems from the University of Strathclyde, Glasgow, U.K., in 1993, 1995, and 2003, respectively. He is currently a Senior Lecturer at the institution. 
His research interests include data engineering and the application of machine learning algorithms to industrial problems including cyber-security. 
\end{IEEEbiography}

\begin{IEEEbiography}
[{\includegraphics[width=1in,height=1.25in,clip,keepaspectratio]{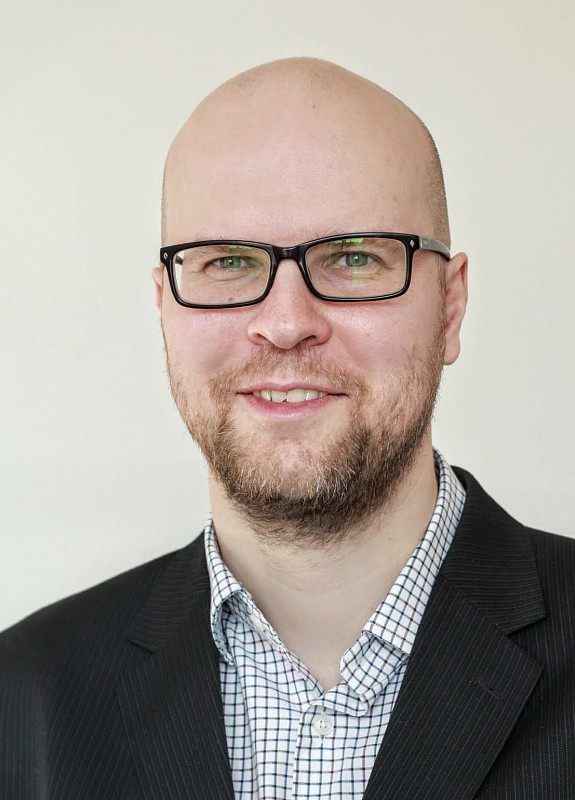}}]{Miroslav Bures} received the Ph.D. degree from the Faculty of Electrical Engineering, Czech Technical University, Prague, where he is currently a Researcher and a Senior Lecturer in software testing and quality assurance. His research interests are model-based testing (process and work flow testing,
data consistency testing) efficiency of test automation, and quality assurance methods for Internet of Things solutions and reflecting specifics of this technology. He is a member of Czech chapter of ACM, CaSTB, and ISTQB Academic work group.
\end{IEEEbiography}

\begin{IEEEbiography}[{\includegraphics[width=1in,height=1.25in,clip,keepaspectratio]{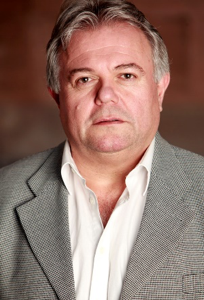}}]{Ivan Andonovic}, BSc, PhD, FIET, SMIEEE, has held a Royal Society Industrial Fellowship in collaboration with British Telecommunications (BT) Labs investigating novel approaches to broadband networking. His current research centres on Internet-of-Things designs and data-driven applications. He has edited two books and authored/co-authored six chapters in books and over 380 journal and conference papers and secured funding for research and development in excess of £10M. He was a member of flagship Scottish Enterprise (Government agency for economic growth) team of the Intermediary Technology Institutes (ITIs), aimed at bridging the gap between basic research and company growth, has been Visiting Scientist at the Communications Research Laboratories of Japan, Visiting Professor at the City University of Hong Kong and Princeton University, Topical Editor for the ‘IEEE Transactions on Communications’ and Technical Programme Co-Chair for the ‘IEEE International Conference in Communications (ICC07)’. 
\end{IEEEbiography}

\begin{IEEEbiography}[{\includegraphics[width=1in,height=1.25in,clip,keepaspectratio]{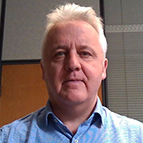}}]{Craig Michie}, BSc, PhD, MIEEE, is the Deputy Head of Department of Electronic and Electrical Engineering at the University of Strathclyde. He obtained a Degree in Electronic and Electrical Engineering and PhD in Coherent Optical Communications from the University of Glasgow. His current research centres on wireless sensor networks, data analytics and applications supported by the Internet-of-Things designs. He has authored/co-authored twelve book chapters, over 200 journal and conference papers and 8 patents.
\end{IEEEbiography}

\begin{IEEEbiographynophoto}{David Brosset} received his master's degree in computer science from the university of south Brittany in 2003 and his Ph.D in computer science in 2008 from the Arts et Metiers, Paris. Since 2011, he is an assistant professor of Computer Science at the French Naval Academy and he is involved in the chair of cyber defence of naval systems. His research concerns the domain of cyber security and in particular the cyber defence of critical systems on board.
\end{IEEEbiographynophoto}

\begin{IEEEbiography}[{\includegraphics[width=1in,height=1.25in,clip,keepaspectratio]{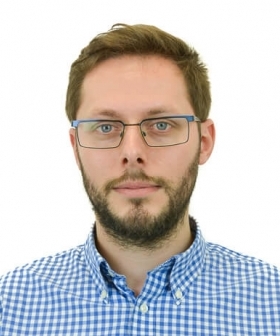}}]{Xavier Bellekens}(S$'$12--M$'$16) received the bachelor’s degree from HeNam, Belgium, in 2010, the master’s degree in ethical hacking and computer security from the University of Abertay, Dundee, in 2012, and the Ph.D. degree in electronic and electrical engineering from the University of Strathclyde, Glasgow, in 2016. He is currently a Chancellor’s Fellow Lecturer with the Department of Electronic and Electrical Engineering, University of Strathclyde, where he has been working on cyber-security for critical infrastructures. Previously, he was a Lecturer in security and privacy with the Department of Cyber-Security, University of Abertay, where he led the Machine Learning for Cyber-Security Research Group. His current research interests include machine learning for cyber-security, autonomous distributed networks, the Internet of Things, and critical infrastructure protection.
\end{IEEEbiography}




\end{document}